\documentclass[12pt]{article}
\usepackage{latexsym}
\usepackage{amssymb}
\catcode`\@=11
\newcount\hour
\newcount\minute
\newtoks\amorpm \hour=\time\divide\hour by 60\minute
=\time{\multiply\hour by 60 \global\advance\minute by-\hour}
\edef\standardtime{{\ifnum\hour<12 \global\amorpm={am}%
        \else\global\amorpm={pm}\advance\hour by-12 \fi
        \ifnum\hour=0 \hour=12 \fi
        \number\hour:\ifnum\minute<10
        0\fi\number\minute\the\amorpm}}
\edef\militarytime{\number\hour:\ifnum\minute<10
0\fi\number\minute}
\def\draftlabel#1{{\@bsphack\if@filesw {\let\thepage\relax
   \xdef\@gtempa{\write\@auxout{\string
      \newlabel{#1}{{\@currentlabel}{\thepage}}}}}\@gtempa
   \if@nobreak \ifvmode\nobreak\fi\fi\fi\@esphack}
        \gdef\@eqnlabel{#1}}
\def\@eqnlabel{}
\def\@vacuum{}
\def\marginnote#1{}
\def\draftmarginnote#1{\marginpar{\raggedright\scriptsize\tt#1}}
\def\draft{
        \pagestyle{plain}
        \overfullrule=2pt
        \oddsidemargin -.1truein
        \def\@oddhead{\sl \phantom{\today\quad\militarytime} \hfil
        \smash{\Large\sl DRAFT} \hfil \today\quad\militarytime}
        \let\@evenhead\@oddhead
        \let\label=\draftlabel
        \let\marginnote=\draftmarginnote
        \def\ps@empty{\let\@mkboth\@gobbletwo
        \def\@oddfoot{\hfil \smash{\Large\sl DRAFT} \hfil}
        \let\@evenfoot\@oddhead}
        \def\@eqnnum{(\theequation)\rlap{\kern\marginparsep\tt\@eqnlabel}%
        \global\let\@eqnlabel\@vacuum}  }

\renewcommand{\theequation}{\thesection.\arabic{equation}}
\renewcommand{\thefootnote}{\fnsymbol{footnote}}
\newcommand{\newsection}{    
\setcounter{equation}{0}\section}
\def\appendix#1{\addtocounter{section}{1}\setcounter{equation}{0}
\renewcommand{\thesection}{\Alph{section}}
\section*{Appendix \thesection\protect\indent \parbox[t]{11.15cm}{#1}}
\addcontentsline{toc}{section}{Appendix \thesection\ \ \ #1}}

\def \lc {{light-cone}}

\def \bi{\bibitem}
\def \la {\label}

\def \b {\beta}
\def \Om {\Omega}

\def \s{\sigma}
\def \d {\partial}

\def\be{\begin{equation}}
\def\ee{\end{equation}}

\def\nat {{\natural}}

\catcode`\@=11
\def \ww {{\rm w} }

\def\bea{\begin{eqnarray}}
\def\eea{\end{eqnarray}}
\def\beann{\begin{eqnarray*}}
\def\eeann{\end{eqnarray*}}
\def\beq{\begin{equation}}
\def\eeq{\end{equation}}
\def\ba{\begin{array}}
\def\ea{\end{array}}
\def\ben{\begin{enumerate}}
\def\een{\end{enumerate}}

\def\s {\sigma }
 \def \la {\label}
 \def\be{\begin{equation}}
\def\ee{\end{equation}}

\def \la {\label}

\font\mybb=msbm10 at 11pt

\def\bb#1{\hbox{\mybb#1}}

\def\bZ {\bb{Z}}
\def\bR {\bb{R}}

\def\bC {\bb{C}}

 \def\ep {\epsilon}

\def \ee {\epsilon}

\def \g {\gamma}
\def \bi{\bibitem}
\def\a{\alpha }

\def \ep {\epsilon}
\def \s {\sigma}

\def \d {\delta}

\def \g {\gamma}

\def \b {\beta}

\def\lc{\lrcorner}

\def\be{\begin{equation}}
\def\ee{\end{equation}}

\def \bi {\bibitem}
\def \la{\label}

\begin{document}
\date{October 2004}
\begin{titlepage}
\begin{center}

\vspace{2.0cm}
{\Large \bf  The spinorial geometry of supersymmetric backgrounds}\\[.2cm]

\vspace{1.5cm}
 {\large J. Gillard, U. Gran and   G. Papadopoulos }

 \vspace{0.5cm}
Department of Mathematics\\
King's College London\\
Strand\\
London WC2R 2LS
\end{center}

\vskip 1.5 cm
\begin{abstract}

We propose a new method to solve the Killing spinor equations of eleven-dimensional
supergravity  based on a description of spinors in terms of forms and on the
$Spin(1,10)$ gauge symmetry of the supercovariant derivative.
 We give the canonical form of Killing  spinors for   backgrounds preserving
 two supersymmetries, $N=2$,  provided
 that one of the spinors  represents the orbit of $Spin(1,10)$ with stability
 subgroup $SU(5)$. We directly solve the Killing spinor equations
 of $N=1$ and some $N=2$, $N=3$ and $N=4$ backgrounds.
 In the $N=2$ case, we investigate backgrounds
 with $SU(5)$ and $SU(4)$ invariant Killing spinors
and compute the associated spacetime forms. We find that $N=2$
backgrounds with $SU(5)$ invariant Killing spinors admit a timelike
Killing vector and that the space transverse to the orbits of this
vector field is a Hermitian manifold with an $SU(5)$-structure.
Furthermore, $N=2$ backgrounds with  $SU(4)$ invariant Killing
spinors admit two Killing vectors, one  timelike and one spacelike.
The space transverse to the orbits of the former is an almost
Hermitian manifold with an $SU(4)$-structure. The spacelike Killing vector field
leaves the almost complex structure invariant.  We explore the canonical form
of Killing spinors for backgrounds preserving more than two
supersymmetries, $N>2$. We investigate a class of  $N=3$ and $N=4$
backgrounds with $SU(4)$ invariant spinors. We find that in both
cases the space transverse to a timelike vector field is a Hermitian
manifold equipped with an $SU(4)$-structure and admits two
holomorphic Killing vector fields. We also present an application to
M-theory Calabi-Yau compactifications with fluxes to one-dimension.

\end{abstract}
\end{titlepage}
\newpage
\setcounter{page}{1}
\renewcommand{\thefootnote}{\arabic{footnote}}
\setcounter{footnote}{0}

\setcounter{section}{0}
\setcounter{subsection}{0}
\newsection{Introduction}
The last ten years, there has been much activity in understanding
the supersymmetric solutions of ten- and eleven-dimensional supergravities.
 This is because of the insight that these solutions give
 in string theory, M-theory and gauge theories, see e.g.
 \cite{townsend, maldacena} and more recently  \cite{gpjf, berenstein, tseytlin}.
Despite these developments, the supersymmetric solutions of eleven-
and ten-dimensional supergravities have not been classified. This is
mainly due to the fact that the supercovariant connections of
supergravity theories in the presence of fluxes are not induced from
connections of the tangent bundle of spacetime. However, progress
has been made in two `extreme' cases. On one end, J.
Figueroa-O'Farrill and  one of the authors classified the maximally
supersymmetric solutions in eleven- and ten-dimensional
supergravities   \cite{jfgpa, jfgpb}. On the other end, J.
Gauntlett, J. Gutowski and S. Pakis  have solved the Killing spinor
equations  of eleven-dimensional supergravity in the presence of one
Killing spinor \cite{pakis, gutowski}.

The main aim of this paper is to propose a new method to solve the Killing
spinor equations of eleven-dimensional supergravity
for any number of Killing spinors. Our method is based on
 the systematic understanding
of spinors that can occur as solutions to the eleven-dimensional
supergravity Killing spinor equations
and the observation that the manifest gauge symmetry of
the eleven-dimensional supercovariant
derivative is $Spin(1,10)$. Because of this, as we shall
explain, the Killing spinors
can be put into a canonical form using $Spin(1,10)$ gauge
transformations. Another ingredient
of the method is the understanding  of the stability
subgroups in $Spin(1,10)$ of any number
of spinors.
For this in the first part of this paper,
we present a description of spinors for an
eleven-dimensional spacetime
in terms of forms by adapting a formalism
developed in \cite{wang} in the context of special holonomy.
This description  simplifies the task of classifying supersymmetric
backgrounds in two ways.

\begin{itemize}
\item First it introduces a basis in the space of spinors which can be used
to {\it directly solve} the Killing spinor equations.

\item Second, it provides a systematic way to find the stability subgroup in $Spin(1,10)$
of $N$ spinors
and to compute the spacetime forms associated with a pair of spinors.
\end{itemize}

The stability  subgroup of  Killing spinors in $Spin(1,10)$
 is a way to characterize (classify)
the Killing spinors for any number of supersymmetries. However,
as we shall see, it is possible that different number of spinors
can have the same stability subgroup. It is well known that there are two
types of orbits, ${\cal O}_{SU(5)}$ and ${\cal O}_{Spin(7)}$, of $Spin(1,10)$
in the space of Majorana spinors, $\Delta_{32}$, with stability
subgroups $SU(5)$ and $(Spin(7)\ltimes \bR^8)\times \bR$,
respectively \cite{bryant, jose}, see also \cite{amus}.
We give  representatives for these two orbits in our
formalism. We then use them to compute
the associated spacetime forms. We shall see that the
relations between the spacetime forms are manifest and there is no need
to use Fierz identities. We would like to point out that another basis
in the space of spinors
has also been  used in \cite{oisin} to directly solve the Killing spinor equations,
 for some $N$,
of a seven-dimensional supergravity\footnote{We thank O. Mac Conamhna for explaining
this result to us.}. This formalism did not employ the description of spinors
in terms of forms that we use. It would be instructive to compare the two methods
for the same supergravity.

Next, we  find the stability subgroups and
 the representatives of orbits for
more than one spinor.
We find that the stability subgroup of two generic spinors,
if one of them has
stability subgroup $SU(5)$, is the identity $\{1\}$. This  has also been observed
for two spinors in seven-dimensions \cite{oisin}.
However there are several special choices of two spinors with
stability subgroups for example $SU(5)$, $SU(4)$, $SU(2)\times SU(3)$, $SU(2)\times SU(2)$,
$Sp(2)$, $SU(3)$
and $SU(2)$. Using the $Spin(1,10)$ gauge symmetry of the supercovariant
connection to put the Killing spinors into
 a canonical form, we give
 the most general expression for the second Killing spinor provided that
the first one represents the orbit ${\cal O}_{SU(5)}$.
We also compute the spacetime forms associated with the spinors with stability
subgroups $SU(5)$ and $SU(4)$.   We shall use these forms to provide a
 geometric characterization
of the associated supersymmetric background.

Using the basis in the space of spinors that we have mentioned, we reduce
the Killing spinor equations of eleven-dimensional supergravity
 to a number of differential and algebraic
conditions which do not contain products of gamma matrices.
To demonstrate the effectiveness
of our formalism, we {\it directly
solve} the Killing spinor equations of eleven-dimensional supergravity
for backgrounds preserving one supersymmetry, $N=1$ backgrounds\footnote
{{}From now on, $N$ is the number of Killing spinors of a supersymmetric
background.},
provided that the Killing spinor
represents the orbit ${\cal O}_{SU(5)}$ of $Spin(1,10)$
in $\Delta_{32}$. The fluxes
are explicitly  related to the spacetime geometry. The spacetime
 admits a timelike Killing vector
field and the space transverse to the orbits of this vector field
is an almost Hermitian manifold.
This is in agreement with the results of \cite{pakis} which have
been derived using a different method.

Next, we focus on the Killing spinor equations for
 backgrounds with $N=2$ supersymmetry.
We solve the Killing spinor equation for the {\it most  general}
$N=2$ background that admits $SU(5)$ invariant Killing spinors. In particular, we
express the fluxes in terms of the geometry of the spacetime.
 We find that the spacetime admits a
time-like Killing vector and that the
manifold transverse to the orbits of the Killing
vector is Hermitian with an $SU(5)$-structure.
We also solve the Killing spinor equations for a class of $N=2$
backgrounds that admit  $SU(4)$ invariant spinors. We find that the spacetime admits a timelike
Killing vector field and a spacelike vector field. The space transverse to the former
is an almost Hermitian manifold with   an $SU(4)$-structure which we determine.
The almost complex structure is invariant under the action of the spacelike vector field.
It is worth pointing out that if the Killing spinors are invariant under a subgroup
$G\subset Spin(10,1)$, then the
spacetime admits a geometric $G$-structure\footnote{Note however that
 one can choose spinors which have  stability
subgroup $\{1\}$. This may limit the applicability of the
 $G$-structure approach for solving the
Killing spinor equations because in such a case any form on
the spacetime is invariant.}.

We also explain how our formalism can be used to classify
$N>2$ supersymmetric backgrounds.
We find that there are many classes of backgrounds with
$N>2$ supersymmetry for
which the spinors have different stability subgroups.
In particular we present an  example of such backgrounds
for which the Killing spinors have stability subgroups
$SU(n)$, $n\leq 5$, which we call the $SU$ series. This class
of backgrounds can be used to investigate M-theory Calabi-Yau compactifications
with fluxes to one, three, five and seven dimensions.

We investigate two classes of backgrounds with $N=3$ and $N=4$ supersymmetry
which admit $SU(4)$ invariant Killing spinors. In both cases,
we solve the Killing spinor equations and express the fluxes in terms of the
geometry of spacetime. We find that the spacetime admits one timelike and two
spacelike Killing vector fields. The space $B$ transverse to the former is a
Hermitian manifold equipped with an $SU(4)$-structure. The two spacelike
Killing vectors are holomorphic on $B$. The space $\hat B$ transverse to all
three Killing vectors is again a Hermitian manifold  with an $SU(4)$-structure.

As an application, we use our results for $N=1$ and $N=2$ backgrounds with
$SU(5)$ invariant Killing spinors  to explore M-theory
Calabi-Yau compactifications with fluxes to one-dimension. We define
Calabi-Yau compactifications with fluxes to one-dimension to be on backgrounds
which are invariant under the Poincar\'e group  of one-dimensional Minkowski space and admit $SU(5)$ invariant Killing
spinors\footnote{One may in addition require
that the internal manifold is compact.}. We find
that such backgrounds can have one or two supersymmetries. We derive the conditions on the spacetime
geometry in both cases. In the latter
case the manifold is a product of the real line and a ten-dimensional
Calabi-Yau manifold. The non-trivial part of the fluxes is given
by a traceless closed (2,2)-form on the Calabi-Yau manifold.

To illustrate the
general method, the supercovariant connection
of eleven-dimensional supergravity \cite{julia} is
\be
{\cal D}_M= \nabla_M + \Sigma_M ~,
\label{supcon}
\ee
where
\be
\nabla_M=\partial_M+{1\over4}\Omega_{M,AB} \Gamma^{AB}
\ee
 is the spin covariant derivative induced from the Levi-Civita connection,
\be
\Sigma_M=-{1\over 288} (\Gamma_M{}^{PQRS}{\rm F}_{PQRS}-8 {\rm F}_{MPQR} \Gamma^{PQR})~,
\label{supconb}
\ee
${\rm F}$ is the four-form field strength (or flux) and $M,N$,
$P,Q$,
$R$,
$S=0, \dots, 9, 10$ are spacetime indices. To find the Killing
spinors is equivalent
to solve the parallel transport problem for the supercovariant
connection. This is related
to the holonomy of the supercovariant connection. It is known that the holonomy
of the supercovariant connection is
$SL(32,\bR)$ \cite{hull, tsimpisa}. At this point, it is crucial to
 distinguish between the
holonomy group $SL(32,\bR)$  and the
gauge group $Spin(1,10)$ of the supercovariant connection.
The latter are the gauge transformations which leave the
form of the supercovariant connection invariant and therefore
are the manifest symmetries of the theory.  Although $SL(32,\bR)$ is the
holonomy of the supercovariant
connection, the $SL(32, \bR)$ gauge transformations, ${\cal D}_M\rightarrow A^{-1} {\cal D}_MA$,
mix the various terms in ${\cal D}_M$  that have different powers of gamma matrices.
As a result, it acts non-trivially of the Levi-Civita connection and the 4-form field
strength ${\rm F}$. On the other hand the $Spin(1,10)$ gauge transformations $U$ give
\be
{\cal D}_M(e, {\rm F})\rightarrow U^{-1} {\cal D}_MU= {\cal D}_M(e', {\rm F}')~,
\ee
where the frame $e$ and the form field strength ${\rm F}$ are related to
$e'$ and ${\rm F}'$ with a local Lorentz rotation.

The existence of Killing spinors is characterized by the reduction of the
holonomy group to subgroups
of  $SL(32,\bR)$ which have been given in \cite{hull, tsimpisa}
 and computed for many supersymmetric
backgrounds in \cite{duffa, duffb}. For a background
with $N$ Killing spinors,  local $SL(32,\bR)$  transformations
can be used to bring them along the first
$N$ vectors in the standard basis of $\Delta_{32}=\bR^{32}$
vector space. This is the `standard' basis
or canonical form
for $N$ Killing spinors up to  local $SL(32,\bR)$ transformations. The simplicity
of this result is facilitated by the property of $SL(32,\bR)$ to have
 one  orbit in $\Delta_{32}-\{0\}$.
However as we have explained such
transformations will not leave the form of the
 supercovariant connection invariant.
Because of this, it is preferable to find the
canonical form of Killing spinors up to
  $Spin(1,10)$ gauge transformations. Because
$Spin(1,10)\subset SL(32, \bR)$, $Spin(1,10)$
has  more orbits in $\Delta_{32}$. In addition, there more subgroups
in $Spin(1,10)$ that preserve $N$ spinors.  As a consequence
there are many more canonical forms for $N$ spinors  up to $Spin(1,10)$
gauge transformations.

Having found the canonical form for $N$ spinors $\{\eta^I: I=1,\dots N\}$
up to $Spin(1,10)$
gauge transformations, one  substitutes them into the Killing spinor
equations.
The resulting equations can be rather involved.
However as we have mentioned, there is a basis
in the space of spinors which is natural within
 our formalism that can be used to reduce
the Killing spinor equations to a set of
differential and algebraic equations which do not involve gamma matrices.
This method works in {\it all cases} and it can be used to
directly solve the Killing spinor equations. In addition  it becomes
 particularly simple and effective for
Killing spinors that have a large stability
subgroup in $Spin(1,10)$.

The paper has been organized as follows:

In section two, we give a description of the spinors, $\Delta_{32}$,
of eleven-dimensional supergravity in terms of forms and explain how
to compute the spacetime forms associated with a pair of spinors. In
section three, we give a representative of the orbit ${\cal
O}_{SU(5)}$ in our formalism and compute the associated spacetime
forms. We also present a basis in the space of spinors which we use
later to analyze the Killing spinor equations. In section
 four, we give the canonical form of the two
spinors that are associated with $N=2$ backgrounds provided that the
first spinor is a representative of the orbit ${\cal O}_{SU(5)}$. We
find that a generic pair of spinors have the identity $\{1\}$ as
stability subgroup in $Spin(1,10)$.
 However there are several special examples with larger stability
subgroups, like $SU(5)$, $SU(4)$ and others. We
give explicitly the spacetime forms
associated with the spinors with stability subgroups $SU(5)$ and $SU(4)$.
In section five, we directly solve the
Killing spinor equations for $N=1$ backgrounds with
a Killing spinor which represents the
orbit ${\cal O}_{SU(5)}$ of the $Spin(1,10)$ gauge
group in $\Delta_{32}$. We also investigate the geometry of the spacetime.
In section six, we solve the Killing spinor
equations for $N=2$ backgrounds for which the
stability subgroup of the Killing spinors is $SU(5)$ and analyze the geometry
of the underlying spacetime.
In section seven, we solve the Killing spinor
equations for $N=2$ backgrounds for which the
stability subgroup of the Killing spinors is $SU(4)$ and analyze the geometry
of the underlying spacetime.
In section eight, we examine the Killing spinors
that can occur in backgrounds with more
than two supersymmetries.
In sections nine and ten, we investigate the geometry of a class
of $N=3$ and $N=4$ backgrounds,
respectively.
In section eleven, we apply our results  to investigate M-theory
 Calabi-Yau compactifications
with fluxes.
In section twelve, we present our conclusions.

In appendix A, we investigate the orbits of $SU(5)$ and $SU(4)$ on
the space of two forms. These orbits are needed to understand the
canonical form of Killing spinors for $N\geq 2$ backgrounds. In
appendix B, we present some aspects of $G$-structures which we use
to analyze the geometry of backgrounds preserving one and two
supersymmetries. In appendix C, we give a representative of the
orbit ${\cal O}_{Spin(7)}$. In appendix D, we present the solution
of the Killing spinor equations for some $N=2$ backgrounds with
$SU(4)$ invariant Killing spinors.

\newsection{Spinors from forms}\label{Spinors}

To find the stability subgroups of spinors  in the
context of eleven-dimensional
supergravity and to simplify many of the computations, we shall use
a characterization of spinors in terms of forms. This has
 been explained for example in \cite{lawson, harvey}
and has been used in \cite{wang} in the context of manifolds with special holonomy.
We shall adapt the construction here
for the spinors of  eleven-dimensional supergravity.
This method can be extended to spinors in all dimensions and all signatures.

A convenient way to describe the Majorana spin representation
$\Delta_{32}$ of $Spin(1,10)$ is to begin from the spin representations
of $Spin(10)$.
Let $V=\bR^{10}$ be a real vector space equipped
with the standard Euclidean inner product.
 The
complex spin (Dirac) representation of $Spin(10)$, $\Delta_c$, is
reducible and decomposes into two irreducible representations,
$\Delta_c=\Delta^+_{16}\oplus \Delta_{16}^-$.

To construct these spin representations  let
$e_1, \dots, e_{10}$ be an orthonormal basis in $V=\bR^{10}$ and $J$ be
a complex structure in $V$, $J(e_i)=e_{i+5}$, $i<6$.  Next consider
the subspace $U=\bR^5$ generated by $e_1, \dots, e_5$. Clearly $V=U\oplus J(U)$.
The Euclidean inner product on $V$ can be extended to a Hermitian inner
product in $V_{\bC}=V\otimes \bC$ denoted by $<,>$, i.e.
\be
<z^i e_i, w^j e_j>=\sum_{i=1}^{10} \bar z^i w^i~,
\ee
where $\bar z^i$ is the standard complex conjugate of $z^i$ in $V_{\bC}$.

The space of Dirac spinors is $\Delta_c=\Lambda^*(U_{\bC})$,
where $U_{\bC}=U\otimes \bC$.
The above inner product can be easily extended to $\Delta_c$ and it is called
the Dirac inner product on the space of spinors.
The gamma matrices  act on $\Delta_c$ as
\bea
\Gamma_i\eta&=& e_i\wedge\eta+e_i\lc \eta~,~~~~ i\leq 5
\cr
\Gamma_{5+i}\eta&=&ie_i\wedge\eta-ie_i\lc \eta~,~~~~ i\leq 5~,
\eea
where $e_i\lc$ is the adjoint of $e_i\wedge$ with respect to $<,>$.
Moreover we have $\Delta_{16}^+=\Lambda^{\rm even}U_{\bC}$ and
$\Delta_{16}^-=\Lambda^{\rm odd}U_{\bC}$.
 Clearly $\Gamma_i: \Delta_{16}^\pm\rightarrow \Delta_{16}^\mp$.
The linear maps $\Gamma_i$ are Hermitian with respect
to the inner product $<,>$, $<\Gamma_i \eta, \theta>=<\eta, \Gamma_i\theta>$,
and satisfy the Clifford algebra relations
$\Gamma_i\Gamma_j+\Gamma_j \Gamma_i=2 \delta_{ij}$.

The charge conjugation matrix is constructed by first  defining the map
$B=\Gamma_6\dots\Gamma_{\nat}$, where\footnote{Form here on, we shall adopt
 the notation to denote
the tenth direction with $\nat=10$.}
$\Gamma_\nat=\Gamma_{10}$. Then  the spinor inner product on $\Delta_c$,
which we denote with the same symbol, is
\be
B(\eta, \theta)=<B(\bar\eta), \theta>~,
\ee
where $\bar\eta$ is the standard complex conjugate of $\eta$ in $\Lambda^*(V_{\bC})$.
It is easy to verify that $B(\eta, \theta)=-B(\theta, \eta)$, i.e. $B$ is skew-symmetric.

It remains to construct $\Gamma_0$ in this representation and impose the Majorana
condition on the spinors. In this case, $\Gamma_0=\pm \Gamma_1\dots \Gamma_{\nat}$
(in what follows we shall choose the plus sign). It is easy to see
that $\Gamma^2_0=-1$ as expected and that $\Gamma_0$ anticommutes with $\Gamma_i$.
The Majorana condition can be easily imposed by setting
\be
\bar\eta= \Gamma_0 B(\eta)~,~~~~~~\eta\in \Delta_c~.
\la{maj}
\ee
The Majorana spinors $\Delta_{32}$ of eleven-dimensional supergravity are those spinors in $\Delta_c$
which obey the Majorana condition (\ref{maj}).
The Pin(10)-invariant inner product $B$ induces a Spin(1,10) invariant inner product
on $\Delta_{32}$ which is the usual skew-symmetric inner product on the space of spinors
of eleven-dimensional supergravity.
This completes the description of  the spinors of eleven-dimensional
supergravity, $\Delta_{32}$, in terms of forms.

One advantage of describing spinor in terms of forms as above
is that it  allows us to  find  the  stability  subgroups  in  $Spin(1,10)$
which leave  certain spinors invariant. It is also useful to bring  spinors
into a canonical or normal form using gauge transformations.
In turn, these simplify the computation of the  space-time forms
\be
\alpha^{IJ}=\alpha(\eta^I, \eta^J)={1\over k!} B(\eta^I,\Gamma_{A_1\dots A_k} \eta^J)
e^{A_1}\wedge\dots\wedge e^{A_k}~,
~~~~~I,J=1,\dots,N~,~~~~~~~k=0,\dots, 9,\nat~
\la{forms}
\ee
associated with spinors.
Note that it is sufficient to compute the forms up to degree five, the rest can be
found using Poincare duality. Because of the symmetry properties of the
gamma matrices and those of the $B$ inner product, $\alpha^{IJ}=\alpha^{JI}$
for forms with degree $k=1,2,5$ and $\alpha^{IJ}=-\alpha^{JI}$ for forms with degree $k=0,3,4$. Therefore,
it is sufficient to compute the spacetime forms with $I\leq J$.

\newsection{$N=1$}

We shall begin with the investigation of the normal form of one spinor
in eleven-dimensional supergravity. As we have mentioned there
 are two orbits ${\cal O}_{SU(5)}$ and ${\cal O}_{Spin(7)}$
of $Spin(1,10)$ in $\Delta_{32}$, one with stability subgroup $SU(5)$ and
the other with stability subgroup $(Spin(7)\ltimes \bR^8)\times \bR$.  We shall mostly
focus on the former case.

\subsection{Spinors with stability group $SU(5)$}

To find the normal form of a spinor up to an $SU(5)\subset Spin(10)\subset
Spin(1,10)$ gauge transformation, we
observe that $\Delta_c$ decomposes under $Spin(10)$ as
\be
\Delta_c=\Delta^+_{16}\oplus \Delta^-_{16}~.
\ee
The representations $\Delta^\pm_{16}$ are complex. The Majorana condition
selects a subspace $\Delta_{32}$ in $\Delta_c$ which intersects both $\Delta^\pm_{16}$.
We have seen that $\Delta^\pm_{16}$ decompose under $SU(5)$ as
\bea
\Delta_{16}^+&=&\sum_{k=0}^2 \Lambda^{2k}(U_{\bC})~,
\cr
\Delta_{16}^-&=&\sum_{k=0}^2 \Lambda^{2k+1}(U_{\bC})~.
\eea
Clearly, the
spinors that are invariant under $SU(5)$ are $1$ and $e_{12345}$. Note that we have adopted
the notation   $e_1\wedge\dots\wedge
e_k=e_{1\dots k}$, e.g. $e_1\wedge e_2\wedge\dots \wedge e_5=e_{12345}$.
Therefore the most general $SU(5)$-invariant spinor  is
\be
\eta=a1+b
e_{12345}~,~~~~~a,b\in \bC~.
\ee
Imposing the Majorana condition on $\eta$,
we find that $ b=\bar a$. Therefore
the $SU(5)$ invariant Majorana spinors are
\be
 \eta=a 1+ \bar a e_{12345}~.
\ee
 So there are {\it two
linearly independent real spinors} invariant under $SU(5)$  given by
\bea
\eta^{SU(5)}&=&{1\over \sqrt{2}} (1+ e_{12345})~,
 \cr
\theta^{SU(5)}&=& {i\over \sqrt{2}}(1-e_{12345})~
\eea
which can represent the orbit ${\cal O}_{SU(5)}$.
Indeed, these two spinors are in the same orbit of $Spin(1,10)$. To see this observe that
\be
\theta^{SU(5)}=\Gamma_0 \eta^{SU(5)}= \Gamma_{1\dots\nat}\eta^{SU(5)}~.
\ee
Therefore the transformation in $Spin(1,10)$ which relates $\eta^{SU(5)}$ with $\theta^{SU(5)}$ projects
onto the Lorentz element which is associated with reflection in all spatial directions.
We shall see later that the forms associated with $\eta^{SU(5)}$ and $\theta^{SU(5)}$ are related by this
Lorentz transformation. Therefore we conclude that in the $SU(5)$ case the parallel spinor $\eta_1$ can always
be chosen as $\eta_1=f \eta^{SU(5)}$, where $f$ is a function of spacetime.

\subsection{Spinors and antiholomorphic forms}

It is convenient for many computations to use another basis in the space of spinors
based on the isomorphism between spinors and $(0,p)$-forms\footnote{On
complex manifolds with an $Spin_c$
structure $\Delta_c=\Lambda^{0,*}\otimes \kappa^{{1\over2}}$,
 where $\kappa$ is the canonical
bundle.}.  In particular it is well known that
\be
\Delta_c=\sum_{k=0}^5\Lambda^{0,k}(\bC^5)~.
\la{isom}
\ee
To see this observe that the $SU(5)$ invariant spinor $1$ satisfies
\be
(\Gamma_j+i \Gamma_{j+5}) 1=0
\ee
and similarly $(\Gamma_j-i \Gamma_{j+5}) e_1\wedge\dots\wedge e_5=0$.
This is the familiar
property of the $SU(5)$ invariant spinors
to be annihilated by the (anti)holomorphic gamma
matrices. In particular,  we define the gamma matrices in a Hermitian basis as
\be
\Gamma_{\bar\a}={1\over \sqrt{2}} (\Gamma_\a+i \Gamma_{\a+5})~,~~~~\a=1,\dots,5
\ee
and $\Gamma^\a=g^{\a\bar\b} \Gamma_\b$, where $g_{\a\bar\b}=\delta_{\a\bar\b}$.
The Clifford algebra relations in this basis are
$\Gamma_\a\Gamma_{\bar \b}+\Gamma_{\bar \b} \Gamma_\a=2g_{\a\bar\b}$ and
$\Gamma_\a \Gamma_\b+\Gamma_\b\Gamma_\a= \Gamma_{\bar\a} \Gamma_{\bar\b}
+\Gamma_{\bar\b}\Gamma_{\bar\a}=0$.
The isomorphism (\ref{isom})
is simply
\be
\Delta_c= \sum_{k=0}^5\Lambda^{0,k} \cdot 1~,
\ee
where $\cdot$ denotes Clifford multiplication. Therefore
\be
\Gamma^{\bar \alpha_1\dots \bar\alpha_k}\cdot 1~,~~~~~k=0, \dots, 5
\la{hbasis}
\ee
is a {\it basis} in the space of spinors $\Delta_c$.
In particular, the other $SU(5)$ invariant spinor can be written as
\be
e_{12345}={1\over 8\cdot 5!} \epsilon_{\bar\a_1\dots\bar\a_5}
 \Gamma^{\bar\a_1\dots\bar\a_5}\cdot 1~,
\ee
where $\epsilon_{\bar 1\bar 2\bar 3\bar 4 \bar 5}=\sqrt{2}$.
We shall extensively use this basis for spinors  to analyze
the Killing spinor equations.

\subsubsection{Forms associated to $\eta^{SU(5)}$}

The spacetime forms (\ref{forms}) associated to the spinor $\eta^{SU(5)}$  are easily
computed. For this first observe that
\bea
B(1,1)&=&B(e_{12345},e_{12345})=0
\cr B(1,e_{12345})&=&-i~.
\eea
Using these we find that the non-vanishing spacetime forms
are the following:

(i) A one-form
\bea
 \kappa=\kappa(\eta^{SU(5)}, \eta^{SU(5)})=B(\eta^{SU(5)}, \Gamma_0\eta^{SU(5)}) e^0=-
e^0~,
\eea
(ii) a two-form
\bea
\omega=\omega(\eta^{SU(5)}, \eta^{SU(5)})=
-e^1\wedge e^{6}-e^2\wedge e^{7}-e^3\wedge e^{8}-e^4\wedge e^{9}-e^5\wedge e^{\nat}
\la{kahlersuf}
\eea
 and
(iii)
a five-form
\bea
\tau(\eta^{SU(5)}, \eta^{SU(5)})={\rm Im}[(e^1+i e^6)\wedge\ldots\wedge(e^5+i
e^\nat)]+\frac{1}{2}e^{0}\wedge \omega\wedge \omega\,.
\eea
All these forms are $SU(5)$ invariant because the associated spinor is $SU(5)$
invariant. The presence of $\omega$ and the first part of $\tau$ may
have been expected because
of the $SU(5)$ invariance. There is also a time-like vector field $\kappa$.
Having found the forms explicitly, it is straightforward to establish their
relations, i.e. $i_{\kappa }\tau={1\over2} \omega\wedge \omega$.
The forms $\kappa, \omega$ and $\tau$ and their relations have also been computed in
\cite{pakis} using different conventions and another method which  involved  Fierz
identities.

\newsection{$N=2$}

Let $\eta_1$ and $\eta_2$ be the Killing spinors of a background
with $N=2$ supersymmetry. It is always possible to choose $\eta_1$
up to an $Spin(1,10)$ gauge transformation to be proportional either
to $\eta^{SU(5)}$ or to $\eta^{Spin(7)}$ ($\eta^{Spin(7)}$ is given
in  appendix C).
 Suppose that $\eta_1= f\eta^{SU(5)}$.
One restriction on the choice of the second Killing spinor $\eta_2$ is that it must
be linearly independent from $\eta_1$ at every spacetime point.  This is because
if two Killing spinors are linearly dependent at one spacetime point,
since the Killing spinor equation
is first order, they will be linearly dependent everywhere and
so they will coincide (up to a constant
overall scale). In addition, it is
sufficient to determine $\eta_2$ up to $SU(5)$ gauge transformations that fix
$\eta_1$.  Using this gauge freedom,  we  decompose $\Delta^+_{16}$
under  $SU(5)$. The second spinor can be chosen as any of the representatives
of the orbits of $SU(5)$ in $\Delta^+_{16}$ which are linearly independent from the
component $1$ of $\eta_1$.
This is sufficient because the Majorana condition determines
the component of the spinor in $\Delta^-_{16}$.

As we have already explained, the $\Delta^+_{16}$ representation of $Spin(10)$
decomposes under $SU(5)$ as
\be
\Delta^+_{16}= \Lambda_{1}^0(\bC^5)\oplus \Lambda_{10}^2(\bC^5)
\oplus \Lambda_{\bar 5}^4(\bC^5)~.
\ee
In this notation the superscript denotes the degree of the forms and the subscript the
dimension of the $SU(5)$ representation.
Since $1\in \Lambda_{1}^0(\bC^5)$, it is clear that the second (Killing)
spinor can be chosen as
\be
\eta_2=b 1+  \theta+{\rm c.c.}~,~~~ b\in \bC
\la{secsp}
\ee
where
$\theta\in  \Lambda_{10}^2(\bC^5)\oplus \Lambda_{\bar 5}^4(\bC^5)$
 and ${\rm c.c}$ denotes
the Majorana complex conjugation. In the
 analysis of the Killing spinor equations the parameter $a$ in
  (\ref{secsp}), as well as the other parameters
  which parameterize the orbits of $SU(5)$,  are promoted
to spacetime functions.

There are four possibilities to choose  $\theta$:

\begin{itemize}

\item $\theta=0$

\item $\theta\in \Lambda_{\bar 5}^4(\bC^5)$,

\item $\theta\in \Lambda_{10}^2(\bC^5)$ and

\item the generic case where
$\theta \in\Lambda_{\bar 5}^4(\bC^5)\oplus\Lambda_{10}^2(\bC^5)$.

\end{itemize}

In some of these cases, there are more than one type of orbit of $SU(5)$. This makes the
description of the choice of $\eta_2$ rather involved. However the analysis can be simplified somewhat
by considering the stability subgroup in $SU(5)$ that leaves invariant both $\eta_1$ and $\eta_2$. Typically,
the cases with `large' stability subgroups are simpler to analyze because
the associated Killing spinors
depend on fewer spacetime functions. Simplifications can also occur by allowing certain components
of the spinor to vanish.

\subsection{$\eta_1=\eta^{SU(5)}$ and $\theta=0$}

  The Killing spinors $\eta_1$ and $\eta_2$ are spanned by the two $SU(5)$
invariant spinors $\eta^{SU(5)}$ and $\theta^{SU(5)}$. Since we can always choose $\eta_1=
a\eta^{SU(5)}$, $a\in \bR$, the second Killing spinor
is
\be
\eta_2=b_1\eta^{SU(5)}+ b_2 \theta^{SU(5)}~,~~~~b_1,b_2\in \bR~.
\la{etasuf}
\ee
The stability subgroup of both $\eta_1$ and $\eta_2$ is $SU(5)$.
Both spinors represent the orbits of $SU(5)$ acting with the trivial
representation in $\Lambda^0(\bC^5)=\bC$. The constants $a,b_1,b_2$ in the context of Killing spinor
equations are promoted to spacetime functions $f, g_1, g_2$, respectively. So generically, the Killing spinors
in this case are determined by three real functions.

Some special cases arise by allowing one or more of $a,b_1,b_2$ to vanish. The first Killing spinor
is chosen such that $\eta_1\not=0$ so $a\not=0$. Thus we have as special cases $b_1=0$ or $b_2=0$. However
 $b_2$ must not vanish, because if $b_2=0$, $\eta_1$ and $\eta_2$ are linearly dependent
  and so they
 coincide. Thus the only special case is that for which $b_1=0$. In this case the
 Killing spinors are $\eta_1=a \eta^{SU(5)}$ and $\eta_2=b \theta^{SU(5)}$, $b_2=b$.

\subsubsection{Forms associated to $\theta^{SU(5)}$}

To compute the spacetime forms associated with $\eta_1$ and $\eta_2$, it is sufficient
to give the forms associated with $\eta^{SU(5)}$ and $\theta^{SU(5)}$. The forms
associated with $\eta_1$ and $\eta_2$ can be computed by taking appropriate linear combinations
of those of $\eta^{SU(5)}$ and $\theta^{SU(5)}$.
The non-vanishing forms associated with $\theta^{SU(5)}$  are
 a one-form
\be
\kappa(\theta^{SU(5)},\theta^{SU(5)})=- e^0~,
\ee
a two-form
\be
\omega(\theta^{SU(5)},\theta^{SU(5)})=-e^1\wedge e^{6}-e^2\wedge e^{7}-e^3\wedge e^{8}-e^4\wedge e^{9}-e^5\wedge e^{\nat}
\la{omm}
\ee
and a five
form
\bea
\tau(\theta^{SU(5)},\theta^{SU(5)})&=&-{\rm Im}[(e^1+i e^6)\wedge\ldots\wedge(e^5+i
e^\nat)]
\cr
&&+\frac{1}{2}e^{0}\wedge \omega(\theta^{SU(5)},\theta^{SU(5)})\wedge \omega(\theta^{SU(5)},\theta^{SU(5)})\,.
\eea
Comparing these with the forms associated with $\eta^{SU(5)}$, we observe
 that $\kappa(\eta^{SU(5)}, \eta^{SU(5)})=\kappa=\kappa(\theta^{SU(5)},\theta^{SU(5)})$ and
$\omega(\eta^{SU(5)}, \eta^{SU(5)})=\omega=\omega(\theta^{SU(5)},\theta^{SU(5)})$
but $\tau(\eta^{SU(5)}, \eta^{SU(5)}) =\tau$ is linearly independent
from $\tau(\theta^{SU(5)},\theta^{SU(5)})$.

\subsubsection{Forms associated to $\eta^{SU(5)}$ and $\theta^{SU(5)}$}

There are also forms associated with the pair of spinors
$(\eta^{SU(5)},\theta^{SU(5)})$. In particular, we have
that there is a zero-form
\be
\alpha(\eta^{SU(5)},\theta^{SU(5)}) =-1
\,,
\ee
 a four-form
\be
\zeta(\eta^{SU(5)},\theta^{SU(5)})=\frac{1}{2}\omega\wedge\omega
\ee
and a five-form
\be
 \tau(\eta^{SU(5)},\theta^{SU(5)})={\rm Re}[(e^1+i e^6)\wedge\ldots\wedge(e^5+i
e^\nat)]\,.
\ee
Therefore the inner product $\alpha$ of $\eta^{SU(5)}$ and
$\theta^{SU(5)}$ is non-degenerate.

\subsection{$\eta_1=\eta^{SU(5)}$ and $\theta\in \Lambda_{\bar 5}^4(\bC^5)$}

There is only one type of orbit  of
$SU(5)$ in $\Lambda_{\bar 5}^4(\bC^5)$ with stability subgroup
$SU(4)$, ${\cal O}_{SU(4)}$. A representative can be chosen as
\be
e_{1234}~.
\ee
Therefore, after imposing the
Majorana condition, we find two real representatives
\be
\eta^{SU(4)}={1\over
\sqrt{2}}(e_5+e_{1234})~,
\ee
and
\be
\theta^{SU(4)}={i\over \sqrt{2}}(e_5- e_{1234})~.
\ee
Observe that
\bea
\eta^{SU(4)}&=&\Gamma_{5} \eta^{SU(5)}~~~~~~~~~\theta^{SU(4)}=
-\Gamma_{5} \theta^{SU(5)}=\Gamma_\nat \eta^{SU(5)}
\cr
\theta^{SU(4)}&=&-\Gamma_0 \eta^{SU(4)}~.
\la{relsuff}
\eea
Therefore the $SU(5)$ and $SU(4)$ invariant spinors are related by an $Spin(1,10)$ transformation
which projects onto the
space reflection $e_1\rightarrow -e_1$, $e_2\rightarrow -e_2$, $e_3\rightarrow -e_3$
and $e_4\rightarrow -e_4$.
In addition, $\eta^{SU(4)}$ and $\theta^{SU(4)}$ related by the $SU(5)$ transformation
\be
e_1\rightarrow e_1~,~~e_2\rightarrow ie_2~,~~e_3\rightarrow ie_3~,~~e_4\rightarrow ie_4~,~~e_5\rightarrow ie_5~.
\ee
 Therefore they represent the same $SU(5)$ orbit.
So in the construction of the second Killing spinor $\eta_2$,
we can choose either $\eta^{SU(4)}$ or $\theta^{SU(4)}$. So the most
general $SU(4)$ invariant
 Killing spinor $\eta_2$ is
\be
\eta_2=b_1 \eta^{SU(5)}+ b_2 \theta^{SU(5)}+b_3 \eta^{SU(4)}~,~~~~~~b_1,b_2,b_3\in \bR~.
\la{skss}
\ee
In this $N=2$ case the Killing spinors depend on four spacetime
functions -- $\eta_1$ depends on one function and $\eta_2$ depends
on three. We also take $b_3\not=0$ because otherwise $\eta_2$ will
reduce to the $SU(5)$ invariant spinor (\ref{etasuf}).
A special case for which the two spinors lie in different orbits is
$\eta_1=a \eta^{SU(5)}$ and $\eta_2=b \eta^{SU(4)}$.

\subsubsection{The spacetime forms of  $\eta^{SU(4)}$}

To compute the forms associated with $\eta_1$ and $\eta_2$ in (\ref{skss}),
 it is sufficient to compute
the spacetime forms associated with $\eta^{SU(4)}$ and the forms associated with the pairs
$(\eta^{SU(5)},\eta^{SU(4)})$ and  $(\theta^{SU(5)},\eta^{SU(4)})$ of spinors.
First let us consider the spacetime forms
$\kappa(\eta^{SU(4)}, \eta^{SU(4)})$, $\omega(\eta^{SU(4)}, \eta^{SU(4)})$ and
 $\tau(\eta^{SU(4)}, \eta^{SU(4)})$
associated with $\eta^{SU(4)}$. Using (\ref{relsuff})
and the forms associated with $\eta^{SU(5)}$,
we find a one-form
\be
\kappa(\eta^{SU(4)}, \eta^{SU(4)})=- e^0~,
\ee
a two-form
\be
\omega(\eta^{SU(4)}, \eta^{SU(4)})=e^1\wedge e^6+e^2\wedge e^7+e^3\wedge e^8+e^4\wedge e^9-e^5\wedge e^\nat~,
\ee
and a five-form
\bea
\tau(\eta^{SU(4)}, \eta^{SU(4)})&=&{\rm Im}[(e^1+i e^6)\wedge\ldots\wedge(-e^5+i
e^\nat)]
\cr
&&+\frac{1}{2}e^{0}\wedge \omega(\eta^{SU(4)}, \eta^{SU(4)})\wedge \omega(\eta^{SU(4)}, \eta^{SU(4)})\,.
\eea

\subsubsection{Forms associated to $(\eta^{SU(5)},\eta^{SU(4)})$
and  $(\theta^{SU(5)},\eta^{SU(4)})$ }

Let us first consider the forms associated with the first pair
$(\eta^{SU(5)},\eta^{SU(4)})$. Using the
relation (\ref{relsuff}) and the forms of $\eta^{SU(5)}$, one
can find that there is a
one-form
\be
\kappa(\eta^{SU(5)},\eta^{SU(4)})=e^\nat\,,
\la{fvf}
\ee
a two-form
\be
\omega(\eta^{SU(5)},\eta^{SU(4)})=-e^0\wedge e^5~,
\ee
a three-form
\be
\xi(\eta^{SU(5)},\eta^{SU(4)})=-\omega^{SU(4)}\wedge e^5~,
\ee
a four-form
\be
\zeta(\eta^{SU(5)},\eta^{SU(4)})={\rm Im}[(e^1+ie^6)\wedge\dots\wedge
(e^4+ie^9)]-e^0\wedge \omega^{SU(4)}\wedge e^\nat
\ee
and a five form
\be
\tau(\eta^{SU(5)},\eta^{SU(4)})=-e^0\wedge {\rm Re} [(e^1+ie^6)\wedge\dots
\wedge (e^4+ie^9)]-{1\over2}\omega^{SU(4)}\wedge
\omega^{SU(4)}\wedge e^\nat~,
\ee
where
\be
\omega^{SU(4)}=e^1\wedge e^{6}+ e^2\wedge e^{7}+e^3\wedge e^{8}+e^4\wedge e^{9}~.
\la{suff}
\ee
Observe that with these spinors there is an associated
spacelike vector $\kappa(\eta^{SU(5)},\eta^{SU(4)})$.
Thus, although with a single spinor one can associate either
a time-like or null vector field, with two or more
spinors one can associate spacelike vectors as well.

The forms associated with the pair of spinors $(\theta^{SU(5)},\eta^{SU(4)})$
can be calculated in a
similar way to find a one-form
\be
\kappa(\theta^{SU(5)},\eta^{SU(4)})= e^5~,
\la{vnff}
\ee
a two-form
\be
\omega(\theta^{SU(5)},\eta^{SU(4)})=e^0\wedge e^\nat~,
\ee
a three-form
\be
\xi(\theta^{SU(5)},\eta^{SU(4)})=\omega^{SU(4)}\wedge e^\nat~,
\ee
a four-form
\be
\zeta(\theta^{SU(5)},\eta^{SU(4)})={\rm Re}[(e^1+i e^6)\wedge\dots
\wedge (e^4+i e^9)]-e^0 \wedge \omega^{SU(4)}
\wedge e^5
\ee
and a five-form
\be
\tau(\theta^{SU(5)},\eta^{SU(4)})=e^0\wedge {\rm Im} [(e^1+ie^6)\wedge\dots
\wedge (e^4+ie^9)]-{1\over2}\omega^{SU(4)}\wedge
\omega^{SU(4)}\wedge e^5~.
\ee

\subsection{$\eta_1=\eta^{SU(5)}$ and $\theta\in \Lambda_{10}^2(\bC^5)$}

There are three different orbits  of $SU(5)$ in
$\Lambda_{10}^2(\bC^5)$ with different
stability subgroups, see appendix A. The generic orbit is
 ${\cal O}_{SU(2)\times SU(2)}$, has stability
subgroup $SU(2)\times SU(2)$  and a representative is
\be
\lambda_1 e_{12}+\lambda_2 e_{34}~,~~~~~~\lambda_1, \lambda_2\in\bR~,
~~~ \lambda_1\not= \lambda_2\not=0~.
\la{tnf}
 \ee
The associated Majorana spinors are
\bea
\eta^{SU(2)\times SU(2)}&=&{1\over \sqrt{2}} (\lambda_1
e_{12}+\lambda_2 e_{34}-\lambda_1 e_{345}- \lambda_2 e_{125})~,
~~~~ \lambda_1^2+\lambda_2^2=1
\cr
\theta^{SU(2)\times SU(2)}&=& {i\over
\sqrt{2}}(\lambda_1 e_{12}+\lambda_2 e_{34}+\lambda_1 e_{345}
+ \lambda_2 e_{125})~.
\eea

In addition there are two special orbits with different stability subgroups.
One special orbit is ${\cal O}_{Sp(2)}$ for $\lambda_1=\pm \lambda_2$
 and so  it has the representative
 \be
 e_{12}\pm e_{34}~.
 \la{sptwo}
\ee
The choice of sign corresponds to different embeddings of
$Sp(2)$ in $SU(5)$. In what follows we choose the
representative with the plus sign. The associated Majorana spinors are
\bea
\eta^{Sp(2)}&=&{1\over {2}} (
e_{12}+ e_{34}-e_{345}-  e_{125})~,
~~~~
\cr
\theta^{Sp(2)}&=& {i\over
{2}}(e_{12}+ e_{34}+e_{345}
+  e_{125})~.
\la{sptspa}
\eea
Incidentally,  in this case there is one more complex $Sp(2)$
invariant spinor which lies in $\Lambda^4_{10}(\bC^5)$
and it is given by wedging (\ref{sptwo}) with itself. The
associated Majorana spinors are
\bea
\zeta^{Sp(2)}&=&{1\over \sqrt{2}} (e_5+e_{1234})
\cr
\mu^{Sp(2)}&=& {i\over \sqrt{2}} (e_5-e_{1234})~.
\la{sptspb}
\eea
The spinors (\ref{sptspa}) and (\ref{sptspb}) together
with $\eta^{SU(5)}$ and $\theta^{SU(5)}$
are the  six  spinors in $\Delta_{32}$ invariant under $Sp(2)$.
It is known that the $Sp(2)$ group
is the holonomy group of eight-dimensional hyper-K\"ahler manifolds.
Supergravity backgrounds
with  $Sp(2)$  holonomy group have been investigated
 before in the context of branes \cite{gppkt, teschen}.

The third special orbit, ${\cal O}_{SU(2)\times SU(3)}$,  arises
whenever $\lambda_1=0$ or $\lambda_2=0$ and has stability subgroup $SU(2)\times SU(3)$.
In the latter case, a representative is
\be
 e_{12}~.
\ee
The associated Majorana spinors are
\bea
\eta^{SU(2)\times SU(3)}&=&{1\over \sqrt{2}} (
e_{12}-e_{345})~,
~~~~
\cr
\theta^{SU(2)\times SU(3)}&=& {i\over
\sqrt{2}}( e_{12}+e_{345})~.
\eea

\subsection{$\eta_1=\eta^{SU(5)}$ and $\theta\in
\Lambda_{\bar 5}^4(\bC^5)\oplus\Lambda_{10}^2(\bC^5)$}

We assume  that the group $SU(5)$ acts non-trivially on both
subspaces of
$\Lambda_{5}^1(\bC^5)\oplus \Lambda_{10}^2(\bC^5)$. If it does not, then this case reduces to cases
investigated in the previous sections. To find the generic orbit of $SU(5)$ in
$\Lambda_{5}^1(\bC^5)\oplus \Lambda_{10}^2(\bC^5)$,
 we choose a generic
element in  $\Lambda_{10}^2(\bC^5)$ and use  the $SU(5)$ transformations to bring the
two-form  in its normal form (\ref{tnf}).
As we have explained, this has stability subgroup $SU(2)\times SU(2)$. This subgroup can then
be used to bring the four-form component of the generic element in
$\Lambda_{5}^1(\bC^5)\oplus \Lambda_{10}^2(\bC^5)$
into a normal form. As a result, a representative
of a generic orbit of $SU(5)$ in $\Lambda_{5}^1(\bC^5)\oplus \Lambda_{10}^2(\bC^5)$  is
\bea
 c_1 e_{2345}+c_2
e_{1245}+c_3 e_{1234}+ b(\lambda_1 e_{12}+\lambda_2
e_{34})~,
 \la{repthree}
\eea
where $
\lambda_1, \lambda_2\in\bR$,  $c_1~,
c_2~, c_3, b \in\bC$, and $\lambda_1\not=
\lambda_2\not=0$ and $c_1\not= c_2\not= c_3\not=0$. To fix a redundancy
in the parametrization, we should
also set $\lambda_1^2+\lambda_2^2=1$. However in what follows  we shall not do so.

The  stability subgroup of $\eta_1=a\eta^{SU(5)}$ and (\ref{repthree}) is  $\{1\}$.
Since the stability subgroup is the identity, this case has the least
residual symmetry. The  complex spinor (\ref{repthree}) will give rise  Majorana spinors
which can be used as Killing spinors for $N=2$ backgrounds.

Apart from the generic orbit with representative (\ref{repthree}), there are various special orbits
which have non-trivial stability subgroups. Since we are interested in the case where
$\theta\in \Lambda_{\bar 5}^4(\bC^5)\oplus\Lambda_{10}^2(\bC^5)$, we shall assume
that at least one of $c_1, c_2, c_3$ and at least one of $\lambda_1, \lambda_2$ do not vanish.

First suppose that $\lambda_1\not=\lambda_2\not=0$.
 If either $c_1$ or $c_2$
vanishes, then the stability
subgroup is $SU(2)$. If both $c_1=c_2=0$, then the stability subgroup is $SU(2)\times SU(2)$.
If $c_1, c_2\not=0$ and $c_3=0$, the stability subgroup is $\{1\}$.

On the other hand if $\lambda_1=\lambda_2$ and either  $c_1$ or $c_2$ vanishes, then
the stability subgroup is $Sp(1)$. The stability subgroup remains the same if in addition $c_3=0$.

Next suppose that either $\lambda_1=0$ or $\lambda_2=0$, say $\lambda_2=0$. If $c_1, c_3\not=0$ and $c_2=0$,
a representative can be chosen as in
(\ref{repthree}) with $c_2=\lambda_2=0$ and the stability subgroup is $SU(2)$. If $c_1=c_2=0$ and $c_3\not=0$,
the stability subgroup is $SU(2)\times SU(2)$ and a representative is given as in (\ref{repthree})
with $c_1=c_2=\lambda_2=0$.  If $c_2=c_3=0$
and $c_1\not=0$, the stability subgroup is $SU(3)$ and a representative is given
by  (\ref{repthree}) with $c_2=c_3=\lambda_2=0$.

\subsection{ The generic case}

 The most general choice of spinors for $N=2$ backgrounds provided one of the spinors
 represents the ${\cal O}_{SU(5)}$ orbit is
\bea
\eta_1&=&a \eta^{SU(5)}
\cr
\eta_2&=&b_1 1+c_1 e_{2345}+c_2
e_{1245}+c_3 e_{1234}+b_2 (\lambda_1 e_{12}+\lambda_2
e_{34})+ {\rm c.c}~,
\eea
where the parameters are as in (\ref{repthree}) and $b_1\in \bC$ and $b_2=b$. For the generic case the stability
subgroup of $\eta_1$ and $\eta_2$  is $\{1\}$. However there are several cases which have larger stability subgroups.
Some examples of $N=2$ spinors  with larger stability subgroups are listed in the table below. Some others have already
been mentioned in the previous sections.
\begin{equation}
\begin{array}{|c|c|}\hline
\mathrm{Conditions \; on\; parameters} & \mathrm{Stability\; subgroups}
 \\
 \hline
c_1=c_2=c_3=\lambda_1=\lambda_2=0\;,\; b_1\not=0& SU(5)\\
    c_2=c_3= \lambda_1=\lambda_2=0\;,\; b_1, c_1\not=0& SU(4) \\
c_1=c_2=c_3= \lambda_2=0\;,\; b_1, \lambda_1\not=0& SU(2)\times SU(3) \\
c_1=c_2=c_3= \lambda_2=0\;,\; b_1, \lambda_1, \lambda_2\not=0\;, \lambda_1\not=\lambda_2& SU(2)\times SU(2) \\
c_1=c_2=c_3= \lambda_2=0\;,\; b_1, \lambda_1, \lambda_2\not=0\;, \lambda_1=\lambda_2& Sp(2)  \\
c_1=c_2= \lambda_2=0\;,\; b_1, \lambda_1, \lambda_2, c_3\not=0\;, \lambda_1\not=\lambda_2& SU(2)\times SU(2) \\
c_1=c_2= \lambda_2=0\;,\; b_1, \lambda_1, \lambda_2, c_3\not=0\;, \lambda_1=\lambda_2& Sp(2)  \\
c_1=c_2= \lambda_2=0\;,\; b_1, \lambda_1, c_3\not=0& SU(2)\times SU(2) \\
c_2=c_3=\lambda_2=0\;,\; b_1, \lambda_1, c_1\not=0&  SU(3) \\
c_1=0\;,\; b_1, \lambda_1, \lambda_2, c_2\not=0&  SU(2)\\
 b_1, \lambda_1,\lambda_2, c_1, c_2\not=0&  \{1\} \\
\mathrm{Generic} & \{1\}\\ \hline
\end{array}
\end{equation}
Some simplification may occur in the form of the spinor $\eta_2$
 using considerations similar to those we have employed
in the $SU(4)$ invariant case to exclude the presence of
 both $\eta^{SU(4)}$ and $\theta^{SU(4)}$ in $\eta_2$.

\newsection{$N=1$ backgrounds}

\subsection{The Killing spinor equations}

To solve the Killing spinor equations, it is convenient to introduce
an orthonormal frame $\{e^A: A=0, \dots, \nat\}$
and to write the spacetime metric as
\be
ds^2= -(e^0)^2+\sum_{i=1}^\nat (e^i)^2~.
\ee
In this frame, the four-form field strength ${\rm F}$ can be expanded
 in electric and magnetic parts as
\be
{\rm F}={1\over 3!} e^0\wedge G_{ijk} e^i\wedge e^j\wedge e^k+ {1\over4!} F_{ijkl}
e^i\wedge e^j\wedge e^k\wedge e^l~.
\ee
The spin (Levi-Civita) connection has non-vanishing components
\be
\Omega_{0, ij}~,~~~~\Omega_{0, 0j}~,~~~~\Omega_{i,0j}~,~~~~\Omega_{i,jk}~.
\ee
In terms of these, the Killing spinor equation decomposes
as
\bea
0&=&\partial_0 \eta+{1\over 4} \Omega_{0, ij}\Gamma^{ij}\eta-{1\over2}\Omega_{0,0i}
\Gamma_0\Gamma^{i}\eta-{1\over 288}\bigl(\Gamma_0 \Gamma^{ijkl}
F_{ijkl}-8 G_{ijk} \Gamma^{ijk}\bigr)\eta\,, \cr
0&=&\partial_i\eta+ {1\over4} \Omega_{i,jk} \Gamma^{jk}\eta-
{1\over2} \Omega_{i,0j} \Gamma_0 \Gamma^j\eta
-{1\over 288} \bigl (\Gamma_i{}^{jklm} F_{jklm}\cr
&&+4 \Gamma_0 \Gamma_i{}^{jkl} G_{jkl}
-24 \Gamma_0 G_{ijk} \Gamma^{jk}- 8 F_{ijkl} \Gamma^{jkl}\bigr) \eta\,.
\eea

The simplification that occurs for $N=1$ backgrounds is that
there is always an $Spin(1,10)$ gauge transformation
to bring the Killing spinors into the form $f \eta^{SU(5)}$ or $f \eta^{Spin(7)}$,
where $f$ is a function
which is restricted by the Killing spinor equations.
We shall focus on the $f \eta^{SU(5)}$ Killing spinor.
 Substituting this into the Killing spinor equations, we have
\bea
\partial_0 \log(f) \eta^{SU(5)}&+&{1\over 4} \Omega_{0, ij}
\Gamma^{ij}\eta^{SU(5)}-{1\over2}\Omega_{0,0i} \Gamma_0\Gamma^{i}\eta^{SU(5)}
\cr
&&-
{1\over 288}\bigl(\Gamma_0 \Gamma^{ijkl} F_{ijkl}-8 G_{ijk}
\Gamma^{ijk}\bigr)\eta^{SU(5)}=0\,,
\cr
\partial_i\log(f) \eta^{SU(5)}&+& {1\over4} \Omega_{i,jk} \Gamma^{jk}\eta^{SU(5)}-
{1\over2} \Omega_{i,0j} \Gamma_0 \Gamma^j\eta^{SU(5)}
-{1\over 288} \bigl (\Gamma_i{}^{jklm} F_{jklm}\cr
&&+4 \Gamma_0 \Gamma_i{}^{jkl} G_{jkl}
-24 \Gamma_0 G_{ijk} \Gamma^{jk}- 8 F_{ijkl} \Gamma^{jkl}\bigr) \eta^{SU(5)}=0\,.
\label{KillingSpinorEqa}
\eea
To analyze these equations, we first observe that they can be
written entirely as ten-dimensional equations
using $\Gamma_0 1=i 1$ and $\Gamma_0 e_{12345}=-i e_{12345}$.
Next, it is convenient to introduce  the Hermitian basis
(\ref{hbasis}) in the space of spinors
 and use the fact that the spinors
$1$ and $e_{12345}$ are annihilated by the holomorphic and
antiholomorphic gamma matrices, respectively, i.e.
$\Gamma^\a 1=0$ and $\Gamma^{\bar\a} e_{12345}=0$.
In addition we write the metric, and decompose the spin connection
 and the fluxes in terms of the Hermitian basis. For example
 the magnetic part of the flux $F$ decomposes as  $(4,0)+(0,4)$,
 $(3,1)+(1,3)$ and $(2,2)$ forms in the usual
 way and similarly the electric part of the flux $G$ decomposes as
 $(3,0)+(0,3)$ and $(2,1)+(1,2)$ forms.
In addition observe that
the spinor $e_{12345}$ can be expressed in terms of the Hermitian basis as
\be
e_{12345}={1\over (\sqrt 2)^5} \Gamma^{\bar 1 \bar 2\dots \bar 5} 1~.
\ee
In this way, we rewrite the Killing spinor equations
 as the vanishing of a spinor expressed in terms of the
 Hermitian basis (\ref{hbasis}).
For such a spinor to vanish
all the components in the Hermitian basis should vanish.
 In particular, the first Killing spinor equation in (\ref{KillingSpinorEqa})
gives
\be
\partial_0 \log f+{1\over2} \Omega_{0,{\a\bar\b}} g^{\a\bar\b}
-{i\over24} F_{\a}{}^{\a}{}_{\b}{}^{\b}=0~,
\la{ksea}
\ee
\be
i \Omega_{0,0 \bar\a}+{1\over3} G_{\bar\a\b}{}^{\b}{}+{i\over 72}
 F_{\b_1\b_2\b_3\b_4} \epsilon^{\b_1\b_2\b_3\b_4}{}_{\bar\a}=0~,
 \la{ksec}
 \ee
 \be
 \Omega_{0,\bar\a\bar\b}-{i\over6} F_{\bar\a\bar\b\g}{}^\g{}
 -{1\over 18} G_{\g_1\g_2\g_3}
 \epsilon^{\g_1\g_2\g_3}{}_{\bar\a\bar\b}=0~.
 \la{ksed}
 \ee
There are three  more conditions arising from the Killing spinor
equations but they are related to the ones
above with (standard) complex conjugation
and therefore are not independent.
The equation (\ref{ksea}) and its complex conjugate imply that
\be
\partial_0\log f=0
\la{kseaa}
\ee
and
\be
 \Omega_{0,{\a\bar\b}} g^{\a\bar\b}={i\over12} F_{\a}{}^{\a}{}_{\b}{}^{\b}~.
 \la{ksebb}
\ee
Therefore $f$ is does not depend on the time frame direction
 and the trace of the magnetic
part of $F$ is determined by the Levi-Civita connection of the spacetime.

The second Killing spinor equation is decomposed in two
parts which involve the derivative of the spinor
along the holomorphic and anti-holomorphic frame directions, respectively.
Since the Killing spinor equation is real, it is sufficient to consider
the Killing spinor
equation with derivatives along the anti-holomorphic frame directions.
The other part involving
derivatives along the holomorphic frame directions is
determined from the anti-holomorphic one
 by standard complex conjugation and therefore it does not give independent equations.
In particular, we find
\be
\partial_{\bar \a} \log f+{1\over2} \Omega_{\bar\a, \b\bar\g} g^{\b\bar\g}
+{i\over12} G_{\bar\a\g}{}^\g
-{1\over 72} \epsilon_{\bar\a}{}^{\b_1\b_2\b_3\b_4} F_{\b_1\b_2\b_3\b_4}=0~,
\la{ksma}
\ee
\be
\partial_{\bar \a} \log f-{1\over2} \Omega_{\bar\a, \b\bar\g} g^{\b\bar\g}
+{i\over4} G_{\bar\a\g}{}^\g=0~,
\la{ksmb}
\ee
\be
i \Omega_{\bar\a,0\bar\b}+{1\over6} F_{\bar\a\bar\b\g}{}^\g
-{i\over 18} \epsilon_{\bar\a\bar\b}{}^{\g_1\g_2\g_3}
G_{\g_1\g_2\g_3}=0~,
\la{ksmc}
\ee
\be
i\Omega_{\bar\a,0\b}+{1\over12} g_{\bar\a\b} F_\g{}^\g{}_\d{}^\d
+{1\over2} F_{\bar\a\b\g}{}^\g=0~,
\la{ksmd}
\ee
\be
\Omega_{\bar\a, \bar\b\bar\g}+{i\over 6} G_{\bar\a\bar\b\bar\g}
-{1\over 12} \epsilon_{\bar\a\bar\b\bar\g}{}^{\g_1\g_2}
F_{\g_1\g_2\d}{}^\d -{1\over 12} F_{\bar\a\g_1\g_2\g_3}
\epsilon^{\g_1\g_2\g_3}{}_{\bar\b\bar\g}=0~,
\la{ksme}
\ee
\be
\Omega_{\bar\a,\b\g}-{i\over 2} G_{\bar\a\b\g}-{i\over3} g_{\bar\a[\b} G_{\g] \d}{}^\d
-{1\over 36} F_{\bar\a\bar\g_1\bar\g_2\bar\g_3}
\epsilon^{\bar\g_1\bar\g_2\bar\g_3}{}_{\b\g}=0~.
\la{ksmf}
\ee

These equations can be viewed either as conditions on the fluxes $F$
and $G$ or conditions on the spacetime geometry as represented by the
Levi-Civita connection $\Omega$. It turns
out that it is convenient to
 express the fluxes in terms of the spacetime
geometry.

\subsection{The solution to the Killing spinor equations}

The above Killing spinor equations can be solved to express the
fluxes in terms of  the Levi-Civita connection of spacetime $\Omega$.
In particular, subtracting (\ref{ksmb}) from (\ref{ksma}), we find that
\be
\Omega_{\bar\a, \b}{}^\b-{i\over6} G_{\bar\a \b}{}^\b-{1\over72}
\epsilon_{\bar\a}{}^{\b_1\dots\b_4} F_{\b_1\dots\b_4}=0~.
\la{ksmaa}
\ee
This equation together with (\ref{ksec}) give
\be
F_{\b_1\dots\b_4}={1\over2} (-\Omega_{0,0\bar\a}+
2 \Omega_{\bar\a, \b}{}^\b) \epsilon^{\bar\a}{}_{\b_1\dots\b_4}
\la{fzF}
\ee
and
\be
G_{\bar\a \b}{}^\b=-2i \Omega_{\bar\a, \b}{}^\b-2i \Omega_{0,0\bar\a}~.
\la{tG}
\ee

Observe that consistency of (\ref{ksmd}) with its complex conjugate requires that
\be
\Omega_{\bar\a,0\b}+\Omega_{\b,0\bar\a}=0~.
\la{fkcon}
\ee
We shall see in the investigation of the geometry that there is a frame such that this condition
can always be satisfied.
Next we take the trace of (\ref{ksmd}) to find
\be
F_{\a}{}^\a{}_\b{}^\b=12 i \Omega_{\bar\a,0\b} g^{\bar\a\b}~.
\la{tF}
\ee
Consistency with (\ref{ksebb}) requires that $\Omega_{0,\b\bar\a} g^{\b\bar\a}+\Omega_{\bar\a,0\b} g^{\bar\a\b}=0$.
We shall see later that this condition is again satisfied with an appropriate choice of frame.
Substituting (\ref{tF}) back into (\ref{ksmd}), we have
\be
F_{\b\bar\a\g}{}^\g=2i \Omega_{\bar\a,0\b}+2i g_{\bar\a\b} \Omega_{\bar\g, 0 \d} g^{\bar\g\d}~.
\la{ttzF}
\ee
Taking the trace of the (\ref{ksmf}) and using (\ref{fzF}) and (\ref{tG}), we get
\be
\Omega_{\bar\a, \b\g} g^{\bar\a\b}-\Omega_{\g,\b}{}^\b-\Omega_{0,0\g}=0~.
\la{gcon}
\ee
This is a condition on the geometry of spacetime which we shall investigate later.
Substituting (\ref{gcon}) back into (\ref{ksmf}), we find that
\be
G_{\bar\a\b\g}=-2i \Omega_{\bar\a,\b\g}+2i g_{\bar\a[\b} \Omega_{0,0\g]}~.
\la{gtco}
\ee

To investigate the equations (\ref{ksmc}) and (\ref{ksme}),
first observe that consistency of (\ref{ksec}) with (\ref{ksmd}) requires that
\be
\Omega_{0, \bar\a\bar\b}=\Omega_{\bar\a, 0\bar\beta}~.
\la{kfconb}
\ee
This again can be satisfied with an appropriate choice of a frame on the spacetime.
These equations can be easily solved
to reveal
\be
G_{\bar\a_1\bar\a_2\bar\a_3}=6i \Omega_{[\bar\a_1,\bar\a_2\bar\a_3]}
\la{gzct}
\ee
and
\be
F_{\bar\a\b_1\b_2\b_3}={1\over2} [\Omega_{\bar\a,\bar\g_1\bar\g_2} \epsilon^{\bar\g_1\bar\g_2}{}_{\b_1\b_2\b_3}
+ 3 \Omega_{\bar\g_1, \bar\g_2\bar\g_3} \epsilon^{\bar\g_1\bar\g_2\bar\g_3}{}_{[\b_1\b_2} g_{\b_3]\bar\a}
+12i \Omega_{[\b_1,0\b_2} g_{\b_3]\bar\a}]~.
\la{otF}
\ee
It remains to solve (\ref{ksma}).  For this substitute (\ref{ksmaa}) back into (\ref{ksma}) and compare it
with (\ref{ksec})   to find
\be
2 \partial_{\bar\alpha} \log f+\Omega_{0,0\bar\a}=0~.
\la{gconb}
\ee

To summarize the solution of the Killing spinor equations,
the electric part of the flux $G$ is completely determined
in terms of the geometry. In particular the (0,3) part, $G^{0,3}$,
is given in (\ref{gzct}) and the (2,1) part, $G^{2,1}$,
is given in (\ref{gtco}).
The rest of the components are determined by standard complex conjugation.
Similarly, the (4,0) and (3,1) parts of the magnetic flux $F$, $F^{4,0}$ and $F^{3,1}$,
 are determined in terms
of the geometry in (\ref{fzF}) and (\ref{otF}), respectively. The
$F^{0,4}$ and $F^{1,3}$ components are also determined from
$F^{4,0}$, and $F^{3,1}$ by standard complex conjugation. In
addition, the trace of $F^{2,2}$ is determined in terms of the
geometry in (\ref{ttzF}).  The Killing spinor equations do not
determine the traceless part of $F^{2,2}$ in terms of the geometry
 and do not involve the traceless part
$\Omega_{i,\b\bar\g}-{1\over5} \Omega_{i, \d}{}^\d g_{\b\bar\g}$
of the connection $\Omega$.

\subsection{The geometry of spacetime}

The one-form $\kappa^f= -f^2 \kappa= f^2 e^0$ is associated with a
 Killing vector field. To see this,
 we have to verify the Killing vector
equation $\nabla_A \kappa^f_B+\nabla_B \kappa^f_A=0$. The
$(A,B)=(0,0)$ component is automatically
satisfied. The $(A,B)=(0, \bar\alpha)$ component is satisfied provided
\be
\partial_{\bar\alpha} f^2 e^0+ \Omega_{0, 0\bar\alpha} f^2 e^0=0
\la{kveqn}
\ee
which is satisfied because of (\ref{gconb}). Similarly, the
Killing vector equations along
$(A,B)=(\a,\b)$ and   $(A,B)=(\a,\bar\b)$ are also satisfied
because of (\ref{fkcon}) and (\ref{kfconb}).

Since $\kappa^f$ is a timelike Killing vector field, one can
always choose coordinates such that $\kappa^f=\partial_t$
and the metric can be written as
\be
ds^2=-f^4 (dt+\alpha)^2+ ds^2_{10}
\ee
where the metric $ds^2_{10}$ on the ten-dimensional space transverse
to the orbits of $\kappa^f$, $f$ and the  one-form $\alpha$ are
independent of $t$. A natural choice of frame is $e^0=f^2
(dt+\alpha)$ with the rest of the components $e^i$ to be  a frame of
$ds^2_{10}$, i.e.
\be
ds^2=- (e^0)^2+\sum^{10}_{i=1} (e^i)^2~.
\ee
Since the frame $e^i$ does not depend on the time coordinate $t$,
 the torsion free condition, $de^i+\Omega^i{}_j \wedge e^j=0$, of the
 Levi-Civita connection of the spacetime for this
frame requires that
\be
\Omega_{i,0j}=\Omega_{0,ij}~.
\ee
As a consequence  the conditions (\ref{fkcon}) and (\ref{kfconb}) are satisfied.

The only remaining condition on the  geometry of the spacetime is
({\ref{gcon}). This restricts the geometry of the ten-dimensional
space $B$ which is transverse to the orbits of the time-like Killing
vector field $\kappa^f$. Because of  (\ref{gconb}), this can be
rewritten as
\be
\Omega_{\bar\a,\b\g} g^{\bar\a\b}-\Omega_{\g,\b}{}^\b+2\partial_\g
f=0~.
\la{gemcon}
\ee
The space $B$ is an almost Hermitian manifold equipped with an $SU(5)$ invariant (5,0)+(0,5)
form $\tau$. Therefore one can use the Gray-Hervella classification of almost Hermitian manifolds
to describe the geometry. In this context,  (\ref{gemcon}) can be written using
appendix B as
\be
(\ww_3)_\g+(\bar\ww_5)_\gamma+2\partial_\g f=0~.
\ee
Equivalently, one can use $(W_5)_i={1\over 40} \epsilon^{j_1\dots
j_5} \nabla_{[i}\epsilon_{j_1\dots j_5]}$ to write\footnote{In the
normalization of $W_5$ we have divided with an additional factor of
$2$ to take into account that in our conventions
$\epsilon_{12345}=\sqrt{2}$.}
\be
 W_5+2df=0~,
\ee
i.e. $W_5$ must be exact. This is the only condition on the geometry
of the almost Hermitian ten-dimensional space $B$ arising from the
Killing spinor equations. The conditions on geometry of the
spacetime that we have described
for $N=1$ are in agreement with those of \cite{pakis}
which have been derived using a different method. Of course there are
more conditions on the
geometry arising from the closure of the fluxes (the Bianchi
identity) and the supergravity field equations.

\newsection{$N=2$ backgrounds with $SU(5)$ invariant Killing spinors}

\subsection{The Killing spinor equations}

As we have explained the most general $SU(5)$ invariant Killing spinors are
\bea
\eta_1&=&f (1+e_{12345})~,
\cr
\eta_2&=& g_1 (1+e_{12345})+ i g_2 (1-e_{12345})~,
\eea
where $f, g_1$ and $g_2$ are real functions on the spacetime which
are restricted by the Killing spinor equations. We shall assume that
$g_2\not=0$ because otherwise the second spinor will be linearly
depend to the first one.

The Killing spinor equation for $\eta_1$ can be analyzed as in the $N=1$ case that we have already explained.
In particular observe that we can write
\be
{\cal D}_M \eta_1= {\cal D}_M [f (1+e_{12345})]=\partial_M f (1+e_{12345})+ f {\cal D}_M (1+e_{12345})=0~
\ee
and thus
\be
{\cal D}_M (1+e_{12345})=- \partial_M\log f (1+e_{12345})~.
\ee
Using this, we can rewrite the Killing spinor equation of $\eta_2$  as
\bea
0&=&{\cal D}_M[g_1 (1+e_{12345})+ i g_2 (1-e_{12345})]
\cr
&=& \partial_M g_1 (1+e_{12345})+ g_1 {\cal D}_M (1+e_{12345})
+{\cal D}_M[i g_2 (1-e_{12345})]
\cr
&=&\partial_M g_1 (1+e_{12345})- g_1 \partial_M\log f (1+e_{12345})
+{\cal D}_M[i g_2 (1-e_{12345})]~.
\la{skseq}
\eea
Multiplying the above equation with $g_2^{-1}$, we find the Killing spinor equation for $\eta_2$ can be expressed as
\be
g_2^{-1}[\partial_M (g_1+i g_2)- g_1 \partial_M\log f]1+
g_2^{-1}[\partial_M (g_1-i g_2)- g_1 \partial_M\log f] e_{12345} +i{\cal D}_M(1-e_{12345})=0~.
\ee
This Killing spinor equation can be analyzed in a way similar to that we
have used for the first Killing spinor. In particular, we first express the equation in terms of ten-dimensional
data and then write  the equation in terms
of the Hermitian basis (\ref{hbasis}). A  difference between the Killing spinor equation of $\eta_2$ (\ref{skseq}) and that
of $\eta_1$ is that the last term has an imaginary unit and there is a relative minus
sign in the terms involving the holomorphic volume form $\epsilon$. Because of this, it is
straightforward to read the conditions arising from Killing spinor equation of $\eta_2$ from those of $\eta_1$.
So we shall not
separately give
the conditions on the connection and the fluxes arising from the second Killing spinor equation. Instead, we shall
proceed to present the analysis of the conditions.

Comparing the Killing spinor equations of $\eta_1$ and $\eta_2$, we find that the Killing spinor
equations that involve the frame time derivative yield the independent equations
\be
\partial_0 f=\partial_0 (g_1+i g_2)=0
\la{timd}
\ee
\be
\Omega_{0,\a\bar\b} g^{\a\bar\b}-{i\over12} F_\a{}^\a{}_\b{}^\b=0
\la{kstea}
\ee
\be
G_{\a_1\a_2\a_3}=F_{\a_1\a_2\a_3\a_4}=0
\la{ksteb}
\ee
\be
i\Omega_{0,0\bar\a}+{1\over 3} G_{\bar\a \b}{}^\b=0
\la{kstec}
\ee
\be
\Omega_{0,\bar\a\bar\b}-{i\over6} F_{\bar\a\bar\b\g}{}^\g=0~.
\la{ksted}
\ee

After a comparison between the Killing spinor equations of $\eta_1$ and $\eta_2$ that involve derivatives along
the spatial directions, we find that
\be
\partial_{\bar\a} \log (g_1 f^{-1})=\partial_{\bar\a} \log (g_2 f^{-1})=0~.
\ee
This together with (\ref{timd}) imply that there are constants $c_1$ and $c_2$ such that
$g_1= c_1 f$ and $g_2=c_2 f$.
Since  Killing spinors are specified up to an overall constant scale, without loss of generality, we introduce an
angle $\varphi$ and  set $g_1=\cos\varphi f, g_2=\sin\varphi f$. The second Killing spinor
can be written\footnote{This
expression is reminiscent of the Killing spinors for dyonic membranes \cite{don}.} as
\be
\eta_2=f[\cos\varphi (1+e_{12345})+ i\sin\varphi (1-e_{12345})]~.
\ee
Therefore $\eta_2$ is determined by the same spacetime function as
$\eta_1$. The angle $\varphi$ is not specified by the Killing spinor
equations. However it is required that $\varphi\not=0,\pi$ because
otherwise $\eta_2$ will not be linearly independent of $\eta_1$.

The rest of the Killing spinor equations for both spinors $\eta_1$ and $\eta_2$ can then be written as
\be
\partial_{\bar \a} \log f+{1\over2} \Omega_{\bar\a, \b\bar\g} g^{\b\bar\g}+{i\over12} G_{\bar\a\g}{}^\g
=0~.
\la{kstma}
\ee
\be
\partial_{\bar \a} \log f-{1\over2} \Omega_{\bar\a, \b\bar\g} g^{\b\bar\g}+{i\over4}
G_{\bar\a\g}{}^\g=0~.
\la{kstmb}
\ee
\be
i \Omega_{\bar\a,0\bar\b}+{1\over6} F_{\bar\a\bar\b\g}{}^\g=0~.
\la{kstmc}
\ee
\be
i\Omega_{\bar\a,0\b}+{1\over12} g_{\bar\a\b}
F_\g{}^\g{}_\d{}^\d+{1\over2} F_{\bar\a\b\g}{}^\g=0~.
\la{kstmd}
\ee
\be
\Omega_{\bar\a, \bar\b\bar\g}=0~.
\la{kstme}
\ee
\be
 \epsilon_{\bar\a\bar\b\bar\g}{}^{\g_1\g_2}
F_{\g_1\g_2\d}{}^\d + F_{\bar\a\g_1\g_2\g_3}
\epsilon^{\g_1\g_2\g_3}{}_{\bar\b\bar\g}=0~.
\la{kstmf}
\ee
\be
\Omega_{\bar\a,\b\g}-{i\over 2} G_{\bar\a\b\g}-{i\over3} g_{\bar\a[\b} G_{\g] \d}{}^\d=0~.
\la{kstmg}
\ee
These equations can be easily solved to reveal the geometry of spacetime.

\subsection{The solution to the Killing spinor equations}

The equation (\ref{ksteb}) implies that the $(4,0)+(0,4)$ and $(3,0)+(0,3)$ parts of $F$ and $G$, respectively, vanish.
The equations (\ref{kstec}) and (\ref{ksted}) can be easily solved to reveal that
\be
G_{\bar\a\b}{}^\b=-3i \Omega_{0,0\bar\a}
\la{kstecc}
\ee
and
\be
F_{\bar\a\bar\b\g}{}^\g=-6i \Omega_{0,\bar\a\bar\b}~.
\la{kstedd}
\ee
Next subtract (\ref{kstmb}) from  (\ref{kstma}), we find that
\be
G_{\bar\a\b}{}^\b=-6i \Omega_{\bar\a, \b}{}^\b~.
\la{kstmbb}
\ee
Comparing this with (\ref{kstecc}), we have the condition
\be
2\Omega_{\bar\a, \b}{}^\b-\Omega_{0,0\bar\a}=0~.
\la{gtcon}
\ee
Substituting (\ref{kstmbb}) into (\ref{kstma}), we find
$\Omega_{\bar\a,\b}{}^\b+\partial_{\bar\a} \log f=0$. Thus
\be
\Omega_{\bar\a,\b}{}^\b={1\over2}\Omega_{0,0\bar\a}=-\partial_{\bar\a} \log f~.
\la{gtcona}
\ee
The solution of the equation (\ref{kstmc}) is
\be
F_{\bar\a\bar\b\g}{}^\g=-6i \Omega_{\bar\a,0\bar\b}
\la{kstmcc}
\ee
and a comparison with (\ref{kstedd}) reveals that
 $\Omega_{\bar\a,0\bar\b}=\Omega_{0,\bar\a\bar\b}$.
We shall show that as in the $N=1$ case, the latter
condition can be satisfied for an appropriate choice of frame.

Consistency of equation (\ref{kstmd}) with its complex conjugate
requires that $\Omega_{\bar\a,0\b}=-\Omega_{\b, 0\bar\a}$. This is
again satisfied for an appropriate choice of frame. Next take the
trace of (\ref{kstmd}) to find
\be
F_\a{}^\a{}_\b{}^\b=12 i\Omega_{\bar\a,0\b} g^{\bar\a\b}~,
\ee
which is compatible with  (\ref{kstea}). Substituting back into (\ref{kstmd}),
we find that
\be
F_{\b\bar\a\g}{}^\g=2i \Omega_{\bar\a,0\b}-2i \Omega_{0, \g}{}^\g g_{\bar\a\b}~.
\la{tfttct}
\ee
The equation (\ref{kstmf}) can be solved  using (\ref{kstmcc}) to give
\be
F_{\bar\a\b_1\b_2\b_3}=6i g_{\bar\a[\b_1} \Omega_{0,\b_2\b_3]}~.
\la{tfoct}
\ee
It remains to investigate equation (\ref{kstmg}). Taking the trace and using (\ref{kstmbb}), we find
that
\be
-\Omega_{\bar\a,\g}{}^{\bar\a} +\Omega_{\g,\b}{}^\b=0~.
\la{gtconb}
\ee
Substituting back into (\ref{kstmg}), we find that
\be
G_{\bar\a\b\g}=-2i \Omega_{\bar \a, \b\g}+2i g_{\bar\a[\b} \Omega_{0,0\g]}~.
\la{tgtcn}
\ee

To summarize, the $(3,0)+(0,3)$ parts of the electric flux $G$ vanish
 and the (2,1) part is determined
in terms of the geometry in (\ref{tgtcn}). The $(4,0)+(0,4)$ parts
of the magnetic flux $F$ vanish and the (1,3) part is determined in
terms of the geometry in (\ref{tfoct}). The trace  of the (2,2) part
of the magnetic flux is determined in terms of the geometry in
(\ref{tfttct}). The traceless part of the  (2,2) component of $F$ is
not specified by the Killing spinor equations. This concludes the
analysis of the solution to the Killing spinor equations.

\subsection{The geometry of spacetime}

Both Killing spinors give rise to the same Killing vector $\kappa^f=f^2 e^0$.
The proof that
$\kappa^f$ is Killing is similar to that in the $N=1$ case and we
 shall not repeat the calculation.
This allows us to adapt coordinates and write the spacetime metric as
\be
ds^2=- f^4 (dt+\alpha)^2+ ds_{10}^2~,
\ee
i.e. as in the $N=1$ case. Consequently, we can find a frame such
that
 $\Omega_{i,0j}=\Omega_{0,ij}$. Thus some of the conditions on the geometry mentioned
in the previous section are satisfied.

The remaining conditions are (\ref{gtcona}), (\ref{gtconb})
and (\ref{kstme}). The latter condition implies that the almost Hermitian
 ten-dimensional manifold $B$ transverse to the orbits
of the Killing vector $\kappa^f$ is complex, i.e. $B$ is Hermitian.
This can be easily seen using the torsion free condition
of the Levi-Civita connection. In particular, we have
\be
de^{\bar\a}=-\Omega_\b{}^{\bar\a}{}_{\bar\g} e^\b\wedge e^{\bar\g}-
\Omega_{\bar\g}{}^{\bar\a}{}_\b  e^{\bar\g}\wedge e^\b-
\Omega_{\bar\b}{}^{\bar\a}{}_{\bar \g} e^{\bar\b}\wedge e^{\bar\g}~.
\ee
and therefore the (2,0) part of $de^{\bar\a}$ vanishes which implies the
integrability of the
complex structure. Equivalently, the condition (\ref{kstme}) implies
the vanishing  of the  Gray-Hervella  classes ${\cal W}_1, {\cal W}_2$, i.e.
\be
{\cal W}_1={\cal W}_2=0~,
\ee
and  $\ww_1=\ww_2=0$.
 It remains to understand  (\ref{gtcona}) and (\ref{gtconb})
 in terms of the Gray-Hervella
classification. Using (\ref{gtconb}), (\ref{gtcona}) can be expressed as
\be
W_5+2 df=0~,
\ee
which is the condition on the geometry that arises for $N=1$
backgrounds. Therefore it remains to explain (\ref{gtconb}). For
this note that
\be
(W_4)_\alpha=2 (\ww_3)_\alpha=2 \Omega_{\bar\b,}{}^{\bar\b}{}_\a~,
\ee
where $(W_4)_i={3\over2} \omega^{jk} \nabla_{[i}\omega_{jk]}$.
Using this, we find that (\ref{gtconb}) can be expressed as
\be
 W_4-W_5=0~.
\la{gtconbb}
\ee
This concludes the discussion of the geometric conditions arising from the
Killing spinor equations with $SU(5)$ invariant  spinors.

\newsection{$N=2$ backgrounds with $SU(4)$ invariant Killing spinors}

\subsection{The Killing spinor equations} \la{sufcon}

The most general $SU(4)$ invariant Killing spinors of $N=2$
backgrounds are
\bea
\eta_1&=&f (1+e_{12345})
\cr
\eta_2&=& g_1 (1+e_{12345})+g_2 i (1-e_{12345})+\sqrt{2} g_3 (e_5+ e_{1234})~.
\eea
where $f$, $g_1, g_2$ and $g_3$ are real functions of the spacetime
which are restricted by the Killing spinor equations. We shall not
investigate the most general case here, this will be presented
elsewhere \cite{prep}. Instead, to simplify the computation we
assume  that $g_1=g_2=0$ and set  $g=g_3\not=0$. The Killing spinor
equation for $\eta_1$ is as in the $N=1$ case. Multiplying the
Killing spinor equation with $g^{-1}$, we find that
\bea
\partial_M\log g (e_5+e_{1234})+
{\cal D}_M (e_5+ e_{1234})=0~.
\eea

To solve the above equation, we first write it in terms of spinors
in ten-dimensions and then use the Hermitian basis (\ref{hbasis}) as
in the previous cases. Then, because of the form of $\eta_2$, the
computation of the conditions arising from the Killing spinor
equations is most easily done by decomposing the fluxes and the
connection in $SU(4)$ representations. In practice this means
splitting the holomorphic  index $\alpha=(\alpha, 5)$, where now
$\alpha$ inside the parenthesis takes values\footnote{Throughout
this section, $\a,\b,\g,\d,\dots=1,2,3,4$~.} $\alpha=1,2,3,4$. Using
this, the conditions arising from Killing spinor equation for
$\eta_2$ involving derivatives along the time direction are
\bea
-i\Omega_{0,05}+{1\over 3} G_{5\a}{}^\a
-{i\over 72} F_{\a_1\a_2\a_3\a_4} \epsilon^{\a_1\a_2\a_3\a_4}=0~,
\la{tfone}
\eea
\bea
\Omega_{0,\bar\a 5}+{i\over 6} F_{\bar\a 5 \b}{}^\b
-{1\over18} G_{\b_1\b_2\b_3} \epsilon^{\b_1\b_2\b_3}{}_{\bar\a}=0~,
\la{tftwo}
\eea
\bea
\partial_0\log g +
{1\over 2} \Omega_{0,\b}{}^\b-{1\over 2} \Omega_{0,5\bar5} +{i\over
24} F_{\a}{}^\a{}_\b{}^\b -{i\over 12} F_\a{}^\a{}_{5\bar 5}=0~,
\la{tfthree}
\eea
\bea
{1\over6} G_{\bar\a\bar\b 5}+ [-{1\over8} \Omega_{0,\g_1\g_2}
-{i\over 48} F_{\g_1\g_2\d}{}^\d
+{i\over 48} F_{\g_1\g_2 5\bar5}]
\epsilon^{\g_1\g_2}{}_{\bar\a\bar\b}=0~,
\la{tffour}
\eea
\bea
{i\over36} F_{\b_1\b_2\b_3 \bar 5} \epsilon^{\b_1\b_2\b_3}{}_{\bar \a}
- {i\over2}
\Omega_{0,0\bar\a}+{1\over6} G_{\bar\a\g}{}^\g-{1\over6} G_{\bar\a5\bar5}=0~.
\la{tffive}
\eea
Similarly, the conditions arising from
 Killing spinor equation for $\eta_2$ involving derivatives
 along the spatial $\bar\a$ directions are
 \bea
-i\Om_{\bar \a,05}+\frac{1}{6}F_{\bar \a5\g}{}^\g
-\frac{i}{18}g_{\bar\a\g_1}G_{\g_2\g_3\g_4}\ep^{\g_1\g_2\g_3\g_4}=0~,
\la{afone}
\eea
\bea
\Om_{\bar\a,\bar\b 5}-\frac{i}{6}G_{\bar\a\bar\b 5}
 -(\frac{1}{12}F_{\bar\a\g_1\g_2\g_3}
    &+&\frac{1}{12}g_{\bar\a\g_1}F_{\g_2\g_3\d}{}^\d
 \cr
 &-&\frac{1}{12}g_{\bar\a\g_1}F_{\g_2\g_3 5\bar 5})\ep^{\g_1\g_2\g_3}{}_{\bar\b}=0~,
 \la{aftwo}
 \eea
\bea
\partial_{\bar \a} \log g
&+&
\frac{1}{2}\Om_{\bar\a,\g}{}^\g-\frac{1}{2}\Om_{\bar\a,5\bar 5}
\cr
   & -&\frac{i}{12}G_{\bar\a\g}{}^\g+\frac{i}{12}G_{\bar\a 5\bar 5}
    -\frac{1}{18}g_{\bar\a\g_1}F_{\g_2\g_3\g_4\bar 5}\ep^{\g_1\g_2\g_3\g_4}=0~,
\la{afthree}
\eea
\bea
\frac{1}{12}F_{\bar\a\bar\b_1\bar\b_25}-\frac{1}{2}(\frac{1}{4}\Om_{\bar\a,\g_1\g_2}
+\frac{i}{8}G_{\bar\a\g_1\g_2}
    +\frac{i}{12}g_{\bar\a\g_1}G_{\g_2\d}{}^\d-\frac{i}{12}g_{\bar\a\g_1}G_{\g_2 5\bar 5})\ep^{\g_1\g_2}{}_{\bar\b_1\bar\b_2}=0~,
\la{affour}
\eea
\bea
-\frac{i}{2}\Om_{\bar\a,0\bar\b}
+\frac{1}{12}F_{\bar\a\bar\b\g}{}^\g
    -\frac{1}{12}F_{\bar\a\bar\b 5\bar 5}+\frac{i}{12}\ep_{\bar\a\bar\b}{}^{\g_1\g_2}G_{\g_1\g_2\bar 5}=0~,
\la{affive}
\eea
\bea
(\frac{i}{2}\Om_{\bar\a,0\g}-\frac{1}{4}F_{\bar\a\g\d}{}^\d
    +\frac{1}{4}F_{\bar\a\g 5\bar 5}-\frac{1}{24}g_{\bar\a\g}F_\d{}^\d{}_\s{}^\s
    +\frac{1}{12}g_{\bar\a\g}F_{5\bar5\d}{}^\d)\ep^{\g}{}_{\bar\b_1\bar\b_2\bar\b_3}=0~,
\la{afsix}
\eea
\bea
\frac{1}{4}\Om_{\bar\a,\bar\b_1\bar\b_2}
-\frac{i}{24}G_{\bar\a\bar\b_1\bar\b_2}-\frac{1}{2}(\frac{1}{8}F_{\bar\a \bar 5\g_1\g_2}
    +\frac{1}{12}g_{\bar\a\g_1}F_{\g_2\bar 5\d}{}^\d)\ep^{\g_1\g_2}{}_{\bar\b_1\bar\b_2}
    =0~,
\la{afseven}
\eea
\bea
\partial_{\bar\a} \log g-\frac{1}{2}\Om_{\bar\a,\g}{}^\g+\frac{1}{2}\Om_{\bar\a,5\bar 5}
    -\frac{i}{4}G_{\bar\a\g}{}^\g+\frac{i}{4}G_{\bar\a 5\bar 5}=0~,
\la{afeight}
\eea
\bea
\frac{1}{72}F_{\bar\a \bar\b_1\bar\b_2\bar\b_3}+\frac{1}{12}(\frac{1}{2}\Om_{\bar\a,\g\bar 5}+\frac{i}{4}G_{\bar\a\g\bar 5}
+\frac{i}{12}g_{\bar\a\g}G_{\bar 5\d}{}^\d)\ep^\g{}_{\bar\b_1\bar\b_2\bar\b_3}=0~,
\la{afnine}
\eea
\bea
 \frac{i}{2}\Om_{\bar\a,0\bar 5}
    -\frac{1}{4}F_{\bar\a\bar 5\g}{}^\g=0~.
 \la{aften}
\eea
The conditions arising from
 Killing spinor equation for $\eta_2$ involving derivatives
 along the spatial $\bar5$ direction are
\bea
i\Om_{\bar
    5,05}+\frac{1}{12}F_\g{}^\g{}_\d{}^\d
    +\frac{1}{3}F_{5\bar 5\g}{}^\g=0~,
    \la{fffone}
\eea
\bea
\Om_{\bar 5,\bar \b 5}-\frac{i}{6}G_{\bar \b\g}{}^\g-\frac{i}{3}G_{\bar \b 5\bar 5}
-{1\over36} F_{\bar5\g_1\g_2\g_3} \ep^{\g_1\g_2\g_3}{}_{\bar\b}=0~,
\la{ffftwo}
\eea
\bea
\partial_{\bar 5} \log g+\frac{1}{2}\Om_{\bar 5,\g}{}^\g
    -\frac{1}{2}\Om_{\bar 5,5\bar 5}-\frac{i}{4}G_{\bar 5\g}{}^\g=0~,
\la{fffthree}
\eea
\bea
\frac{1}{12}F_{\bar\b_1\bar\b_2\g}{}^\g+\frac{1}{6}F_{\bar\b_1\bar\b_2 5\bar 5}+
\frac{1}{2}(\frac{1}{4}\Om_{\bar 5,\g_1\g_2}+\frac{i}{24}G_{\bar 5\g_1\g_2})\ep^{\g_1\g_2}{}_{\bar \b_1\bar \b_2}=0~,
\la{ffffour}
\eea
\bea
-\frac{i}{2}\Om_{\bar 5,0\bar\b}-\frac{1}{4}F_{\bar\b\bar 5\g}{}^\g=0~,
 \la{fffive}
\eea
\bea
-\frac{i}{36}G_{\bar\b_1\bar\b_2\bar\b_3}+
\frac{1}{12}(\frac{i}{2}\Om_{\bar 5,0\g}
    -\frac{1}{12}F_{\bar 5\g\d}{}^\d)\ep^{\g}{}_{\bar \b_1\bar \b_2\bar \b_3}=0~,
 \la{fffsix}
\eea
\bea
\Om_{\bar 5,\bar \b_1\bar \b_2}
    -\frac{i}{2}G_{\bar 5\bar \b_1\bar \b_2}=0~,
\la{fffseven}
\eea
\bea
-\frac{1}{144}F_{\bar\b_1\bar\b_2\bar\b_3\bar\b_4}+\frac{1}{96}(\partial_{\bar 5} \log g
    -\frac{1}{2}\Om_{\bar 5,\g}{}^\g+\frac{1}{2}\Om_{\bar 5,5\bar 5}
    -\frac{i}{12}G_{\bar 5\g}{}^\g)\ep_{\bar \b_1\bar \b_2\bar \b_3\bar \b_4}=0~,
\la{fffeight}
\eea
\bea
-F_{\bar\b_1\bar\b_2\bar\b_3\bar 5}+\Omega_{\bar 5,\gamma\bar 5}
\epsilon^\g{}_{\bar\b_1\bar\b_2\bar\b_3}=0~.
\la{fffnine}
\eea

\subsection{Solution to the Killing spinor equations} \la{sumsum}

To solve the Killing spinor equations for $\eta_2$, we use the
solutions to the Killing spinor equations for $\eta_1$. The latter
relate certain components of the fluxes to  spacetime geometry as
represented by the components of the spin connection. We  substitute
all these relations into the conditions that arise from the Killing
spinor equations for  $\eta_2$. As a result, we  turn the conditions
associated with the Killing spinor equations for $\eta_2$ into
conditions on the geometry of spacetime. The calculation is long but
routine and it is explained in detail in appendix D. Here we
summarize the conditions that arise from the analysis.

The Killing spinor equations imply that
\be
g=f~,~~~\partial_0f=(\partial_{5}-\partial_{\bar 5})f=0~.
\la{summary00}
\ee
In fact $g$ is proportional to $f$ but as we have mentioned Killing spinors are
determined up to a constant scale.
The conditions on the $\Omega_{0,0i}$ components are
\be
\Omega_{0,05}=\Omega_{0,0\bar 5}=-2\partial_5 \log f=-2 \partial_{\bar5} \log f~,~~~~
\Omega_{0,0\alpha}=-2\partial_{\a} \log f~.
\la{summary1}
\ee
The conditions on the  $\Omega_{0,ij}$ components  are
\be
\Omega_{0, 5\bar\a}=\Omega_{0, 5\a}= \Omega_{0, 5\bar5}=\Omega_{0,\b}{}^\b=0~,~~~~~
\Omega_{0, \b_1\b_2}={i\over4} (\Omega_{5,\bar \g_1\bar\g_2}-
\Omega_{\bar 5,\bar \g_1\bar\g_2})\epsilon^{\bar\g_1\bar\g_2}{}_{\b_1\b_2}
\la{summary2}
\ee
and the traceless part of $\Omega_{0, \a\bar\b}$ is not determined.
The conditions on the   $\Omega_{\bar\a, ij}$ components are
\bea
\Omega_{[\bar\b_1,\bar\b_2\bar\b_3]}&=&0~,~~~
\Omega_{\bar\a, \b_1\b_2}=-\Omega_{0,0[\b_1} g_{\b_2]\bar\a}~,~~~
\Omega_{ \b,\bar\a}{}^\b={3\over2} (\Omega_{\bar5, \bar\a \bar 5}-\Omega_{\bar5, \bar\a  5})
=-{3\over 2} \Omega_{0,0\bar\a}~,
\cr
\Omega_{\a,\b}{}^\b&=&-{1\over2}(\Omega_{\bar5, \a \bar 5}+\Omega_{\bar5, \a  5})
=-{1\over2} (\Omega_{0,0\a}+2\Omega_{5,\a5})~.
\la{summary3}
\eea
In addition, we have
\bea
&&\Omega_{[\bar\b_1, \bar\b_2]\bar 5}=-\Omega_{\bar 5,\bar\b_1 \bar\b_2}~,~~~
\Omega_{[\bar\b_1, \bar\b_2] 5}=-\Omega_{5,\bar\b_1 \bar\b_2}~,~~~
\Omega_{(\bar\b_1, \bar\b_2) 5}=\Omega_{(\bar\b_1, \bar\b_2) \bar 5}~,
\cr
&&\Omega_{(\bar\a, \b) \bar5}=\Omega_{(\bar\a, \b) 5}={1\over2} g_{\bar\a\b} \Omega_{0,0\bar 5}~,~~~
\Omega_{\bar\a, 5\bar5}=0~.
\la{summary4}
\eea
Finally, the conditions on the  $\Omega_{\bar5, ij}$ components are
\bea
&&\Omega_{5,\b}{}^\b=\Omega_{\bar5,\b}{}^\b~,~~~
 \Omega_{ 5, \bar\a 5}=\Omega_{\bar 5, \bar\a \bar5}~,~~~
\Omega_{ 5, \bar\a \bar5}=\Omega_{\bar 5, \bar\a 5}~,
\cr
&&\Omega_{\bar5, \bar\a \bar 5}-\Omega_{\bar5, \bar\a  5}= -\Omega_{0,0\bar\a}~,~~~
\Omega_{\bar5, 5\bar 5}=-\Omega_{5,5\bar 5}=-\Omega_{0,0\bar 5}~.
\la{summary5}
\eea
The above equations  together with their complex conjugates give full set of conditions
 that are required for a background
to admit $N=2$ supersymmetry and Killing spinors given by $\eta_1=f \eta^{SU(5)}$ and
$\eta_2=g \eta^{SU(5)}$. The traceless part of $\Omega_{\a, \b\bar\g}$ is not determined
by the Killing spinor equations.

\subsection{Fluxes}

The conditions we have derived for the spin  connection in turn restrict the form of the fluxes. In particular
we find that the electric part of the flux is
\bea
&&G_{\a\b\g}=0~,~~~~G_{5\b\g}=2i \Omega_{5,\b\g}~,~~~G_{\bar\a 5\g}=-2i \Omega_{\bar\a,5\g}-i
g_{\bar\a\g} \Omega_{0,05}~,
\cr
&&G_{\bar5 \b\g}=-2i \Omega_{\bar5,\b\g}~,~~~G_{\bar\a \b\g}=0~,~~~G_{\bar 5 5\a}=-2i \Omega_{\bar5,5\a}+i
\Omega_{0,0\a}
\eea
Similarly, the magnetic part of the flux is
\bea
&&F_{\a_1\a_2\a_3\a_4}={1\over2} (-3\Omega_{0,05}+2\Omega_{5,\b}{}^\b) \epsilon_{\a_1\a_2\a_3\a_4}
~,~~~F_{5\a_1\a_2\a_2}={1\over2} (\Omega_{0,0\bar\b}-2 \Omega_{\bar\b,\g}{}^\g) \epsilon^{\bar\b}{}_{\a_1\a_2\a_3}
\cr
&&F_{\bar\a\b_1\b_2\b_3}={1\over2} [2 \Omega_{\bar\a, \bar5\bar\g}
\epsilon^{\bar\g}{}_{\b_1\b_2\b_3}-3 \Omega_{5,\bar\g_1\bar\g_2}
\epsilon^{\bar\g_1\bar\g_2}{}_{[\b_1\b_2} g_{\b_3]\bar\a}]~,~~~
F_{\bar 5\b_1\b_2\b_3}=-\Omega_{\bar5,\bar\g\bar5} \epsilon^{\bar
\g}{}_{\b_1\b_2\b_3}
\cr
&& F_{\bar\a 5\b_1\b_2}={1\over2} \Omega_{\bar\a, \bar\g_1\bar \g_2}
\epsilon^{\bar\g_1\bar \g_2}{}_{\b_1\b_2} ~,~~~F_{5\bar5
\a}{}^\a=0~,~~~F_{\a\bar\b\g}{}^\g=0
\cr
&& F_{\a\b5\bar5}={1\over2}
(\Omega_{5,\bar\g_1\bar\g_2}-\Omega_{\bar 5,\bar\g_1\bar\g_2})
\epsilon^{\bar\g_1\bar \g_2}{}_{\a\b}~,~~~ F_{\a\bar\b 5\bar5}=-2i
\Omega_{0, \a\bar\b}~,~~~
\cr
&&
F_{\bar \a\bar5 \b_1\b_2}={1\over2}
\Omega_{\bar\a,\bar\g_1\bar\g_2} \epsilon^{\bar\g_1\bar\g_2}{}_{\b_1\b_2}
\eea
The last two relations are derived from the conditions for $N=2$
supersymmetry (\ref{solafour}) and (\ref{solafive}) in appendix D,
respectively. The components of the fluxes that do not appear in the
above equations are not determined by the Killing spinor equations.

\subsection{The geometry of spacetime}

We shall now investigate some aspects of the spacetime geometry that
arises from the  relations (\ref{summary00})-(\ref{summary5}). The
spacetime admits a timelike Killing vector field $\kappa^f$ which is
inherited from the Killing spinor $\eta_1$ as in the $N=1$ case.
From the forms associated with the  spinors $\eta^{SU(5)}$ and
$\eta^{SU(4)}$, it is clear that the spacetime has an
$SU(4)$-structure. In particular, the space $B$  transverse to the
orbits of $\kappa^f$ has an $SU(4)$-structure. The space $B$ is not
complex because the almost complex structure associated with
$\omega(\eta^{SU(5)}, \eta^{SU(5)})$ is {\it not integrable} as can
be seen by looking at the components of the spin connection. In
particular the component $\Omega_{\bar5,\bar\a\bar 5}$ of the connection is
not required to vanish by the Killing spinor equations.  This is a
difference between the $N=2$ backgrounds  with $SU(5)$ and $SU(4)$
invariant structures. In the former case, $B$ is complex.

As we have seen, there is a one-form (\ref{fvf}) constructed from
the spinors $\eta^{SU(5)}$ and  $\eta^{SU(4)}$. Using this, we can
define the vector field
\be
\tilde\kappa^f=f^2\partial_\nat=if^2
(\partial_{5}-\partial_{\bar5})~.
\ee
It turns out that this is
a Killing vector field on the spacetime.
 To show
 this, we have to show that $\tilde\kappa^f$
solves the Killing vector equation,
$\nabla_A\tilde\kappa^f_B+\nabla_B\tilde\kappa^f_A=0$.
In terms of the connection, this can be written as
\be
\partial_A f^2 (\tilde\kappa)_B+ \partial_B f^2 (\tilde\kappa)_A-
\Omega_{A,}{}^C{}_B  (\tilde\kappa^f)_C
- \Omega_{B,}{}^C{}_A  (\tilde\kappa^f)_C=0~,
\ee
where the non-vanishing components of the associated one-form to the
vector field are $(\tilde \kappa)_5=- (\tilde \kappa)_{\bar
5}=-i$ and $\tilde\kappa^f=f^2 \tilde \kappa$. This equation can be easily verified using the equations
(\ref{summary00})-(\ref{summary5}).  For example setting
$(A,B)=(0,0)$ we get, by using that $f$ is time-independent
(\ref{summary00}), that
\be
2i\Omega_{0,}{}^5{}_0
-2i\Omega_{0,}{}^{\bar 5}{}_0 =0
\ee
which vanishes due to (\ref{summary1}). In a similar way, one can show that
$\tilde \kappa^f$ is a Killing vector using the rest of the conditions.

In fact $\tilde \kappa^f$ preserves the almost complex structure as
well, i.e. ${\cal L}_{\tilde
\kappa^f}\omega(\eta^{SU(4)},\eta^{SU(4)})=0$. The computation can
be simplified by expressing the Lie derivative in terms of the spin
connection as
\be
({\cal L}_{\tilde \kappa^f}\omega)_{AB}=2\partial_{[A}
f^2 \omega_{|C| B]} (\tilde\kappa)^C-
2 (\tilde\kappa^f)^D \Omega_{[D,}{}^C{}_{A]} \omega_{CB}+
 2 (\tilde\kappa^f)^D \Omega_{[D,}{}^C{}_{B]} \omega_{CA}
\ee
where $\omega=\omega(\eta^{SU(5)},\eta^{SU(5)})$. Then one can
proceed to verify the equation ${\cal L}_{\tilde \kappa^f}\omega=0$
using the condition (\ref{summary00})-(\ref{summary5}). In fact
$\kappa^f$ also preserves the almost complex structure of $B$. It is
likely that both $\kappa^f$ and $\tilde \kappa^f$ preserve the whole
of the $SU(4)$ structure of $B$ including the higher degree forms
that are associated with the spinors $\eta^{SU(5)}$ and
$\eta^{SU(4)}$.

In addition, one can show that $[\kappa^f,\tilde \kappa^f]=0$. This
is because all the components of the connection of the type
$\Omega_{0,5 i}=0$ and also due to (\ref{summary00}). In such a
case, one can introduce coordinates $u^a$ adapted to both Killing
vector fields and write the metric as
\be
ds^2=U_{ab} (du^a+\b^a) (du^b+\b^b)+ \gamma_{IJ} dx^I dx^J~,
\ee
where $U_{ab}$, $\b$ and $\gamma$ depend on the remaining coordinates $x^I$.

To summarize, we have shown that the $N=2$ backgrounds with  $SU(4)$
invariant spinors admit  two commuting Killing vector fields
$\kappa^f$ and $\tilde \kappa^f$. The former is timelike while the
latter is spacelike. The space $B$ transverse to the orbits of
$\kappa^f$ is an almost Hermitian manifold with an $SU(4)$-structure
with a Killing vector field which preserves the almost complex
structure. The $SU(4)$-structure on $B$ is determined by the
conditions (\ref{summary1})-(\ref{summary5}), see also appendix B.
Of course these conditions can be put into real form using the
almost complex structure and the forms associated with the spinors.

\newsection{$N>2$ backgrounds}

\subsection{General case}

To investigate backgrounds with more than two supersymmetries, one
can repeat the procedure that we have used for the $N=2$ case.
Suppose that we have chosen the first two Killing spinors to be in
the directions $\eta_1$ and $\eta_2$ with $\eta_1$ representing the
orbit ${\cal O}_{SU(5)}$. To choose the third Killing spinor, we
decompose $\Delta_{16}^+$ under the action of the stability subgroup
$H\subset SU(5)$ that leaves both $\eta_1$ and $\eta_2$ invariant.
Typically $H$ is an $SU(n)$ or a product of $SU(n)$ groups. Then, we
look at the orbits of $H$ in $\Delta_{16}^+$. Using the results in
appendix A, it is straightforward to find these orbits. The third
Killing spinor $\eta_3$ can be chosen as a linear combination of the
representatives of these orbits and linearly independent of $\eta_1$
and $\eta_2$.

Clearly this procedure can be repeated to find representatives for
any number of Killing spinors. This method works well in the cases
that the stability subgroup of the spinors is large because it can
be used to restrict the choice of the next spinor and to solve  the
Killing spinor equations. In the case that the stability subgroup is
small,  further progress depends on the details of the Killing
spinor equation. For example, suppose that $\eta_1$ and $\eta_2$ are
chosen such that the stability subgroup is $\{1\}$. If this is the
case, the third spinor can be chosen as any other spinor which is
linearly independent of $\eta_1$ and $\eta_2$. Although our
formalism can still be used, there is no apparent simplification in
the computation of the Killing spinor equations and the forms
associated with $\eta_3$. As a result, the geometry of the
background will be rather involved. Because of this, we shall focus
on those $N>2$ backgrounds which admit spinors with large symmetry
groups. Amongst the various cases, there is one for which
 the  spinors  are invariant under $SU$ groups. To illustrate
 the general procedure of constructing
 canonical forms for the Killing spinors outlined
 above, we will analyze this case further below.

\subsection{The $SU$ series}

The $SU$ series is characterized by the property that the Killing spinors
are progressively invariant under $1\subset SU(2)\subset SU(3)\subset SU(4)\subset SU(5)$.
This series can also be thought
of as the Calabi-Yau series. The Killing spinors that we give below are those expected
in M-theory Calabi-Yau compactifications (with fluxes).

We begin by choosing $\eta_1=\eta^{SU(5)}$ and
$\eta_2=\theta^{SU(5)}$\footnote{We could have taken $\eta_2$
to be a linear combination of $\eta^{SU(5)}$ and $\theta^{SU(5)}$ without
changing the analysis
below but for simplicity we have set  $\eta_2=\theta^{SU(5)}$.} which
 are invariant under $SU(5)$. There are no other
linearly independent spinors invariant under $SU(5)$. To find the
spinors invariant under $SU(4)\subset SU(5)$, recall that we have
analyzed all possible choices of spinors  which are linearly
independent from $\eta_1$ and $\eta_2$. We have found that the only
$SU(4)$ invariant spinors are $\eta^{SU(4)}$ and $\theta^{SU(4)}$.
Therefore we choose\footnote{Again, for $\eta_3$ and $\eta_4$ we
could have chosen a linear combination of $SU(4)$ and $SU(5)$
invariant spinors but for simplicity we have not done so. Similar
considerations arise below for spinors invariant under other $SU$
groups.}
\be
\eta_3=\eta^{SU(4)}~,
\ee
and
\be
\eta_4=\theta^{SU(4)}~.
\ee
The stability subgroup of all the spinors $\eta_1, \dots, \eta_4$ is $SU(4)$. Therefore there are four $SU(4)$
invariant spinors  which is the same number of spinors as those expected for a compactification
of M-theory on an eight-dimensional Calabi-Yau manifold. To proceed,
we decompose $\Delta^+_{16}$ under $SU(4)$. Of course $\Lambda^0(\bC^5)$ does
not decompose further. The one and three
forms decompose as
\bea
\Lambda^1_{5}(\bC^5)=\Lambda^0(\bC^5)\oplus \Lambda_{4}^1(\bC^4)~,
\cr
\Lambda^3_{10}=\Lambda_{6}^2(\bC^4)\oplus \Lambda_{4}^3(\bC^4)~.
\eea
Using the results of appendix A, it is easy to see that the only additional
$SU(3)$ invariant spinors
are
\be
e_4~,~~~~~~~e_{123}~.
\ee
Each of these gives two Majorana spinors. Therefore, we find four additional
spinors which are invariant under $SU(3)$.
These are
\bea
\eta_1^{SU(3)}&=&{1\over\sqrt{2}} (e_4-e_{1235})~,~~~~~
\theta_1^{SU(3)}={i\over\sqrt{2}} (e_4+e_{1235})~,
\cr
\eta_2^{SU(3)}&=&{1\over\sqrt{2}} (e_{45}-e_{123})~,~~~~~
\theta_2^{SU(3)}={i\over\sqrt{2}} (e_{45}+e_{123})~.
\eea
Therefore there are eight $SU(3)$ invariant spinors, i.e. as many as
the number of Killing spinors expected for a compactification of
M-theory on a six-dimensional Calabi-Yau manifold.

To proceed, we decompose $\Lambda_{4}^1(\bC^4)$, $\Lambda_{6}^2(\bC^4)$ and $\Lambda_{4}^3(\bC^4)$ under $SU(3)$
to find
\bea
&&\Lambda^1_{4}(\bC^4)=\Lambda^0(\bC^3)\oplus \Lambda_{3}^1(\bC^3)~,
\cr
&&\Lambda^3_{4}(\bC^4)=\Lambda^2(\bC^3)\oplus \Lambda_{1}^3(\bC^3)~,
\cr
&&\Lambda^2_{6}(\bC^4)=\Lambda_{3}^1(\bC^3)\oplus
\Lambda_{3}^2(\bC^3)~.
\eea
It is easy to see that the additional $SU(2)$ invariant complex spinors are
\be
e_3~,~~~~e_{345}~,~~~~e_{124}~,~~~~e_{125}
\ee
Each of these complex spinors gives rise to two Majorana spinors which are given by
\bea
\eta_1^{SU(2)}&=&{1\over\sqrt{2}} (e_3+e_{1245})~,~~~~~
\theta_1^{SU(2)}={i\over\sqrt{2}} (e_{3}-e_{1245})~,
\cr
\eta_2^{SU(2)}&=&{1\over\sqrt{2}} (e_{12}-e_{345})~,~~~~~
\theta_2^{SU(2)}={i\over\sqrt{2}} (e_{12}+e_{345})~,
\cr
\eta_3^{SU(2)}&=&{1\over\sqrt{2}} (e_{35}+e_{124})~,~~~~~
\theta_3^{SU(2)}={i\over\sqrt{2}} (e_{35}-e_{124})~,
\cr
\eta_4^{SU(2)}&=&{1\over\sqrt{2}} (e_{34}-e_{125})~,~~~~~
\theta_4^{SU(2)}={i\over\sqrt{2}} (e_{34}+e_{125})~.
\eea
The total number of $SU(2)$ invariant spinors is sixteen, i.e. as
many as those expected for an M-theory compactification on $K_3$.
Giving here all the spinors explicitly allows one to construct all
the forms associated with these spinors and in this way get an
insight into the geometry of the supersymmetric background. We shall
not present the results of the Killing spinor analysis here. These
can be found elsewhere \cite{prep}.

\newsection{$N=3$ backgrounds with $SU(4)$ invariant  spinors}

To investigate a class of $N=3$ backgrounds it suffices to combine
the conditions we have derived for $N=2$ backgrounds with $SU(5)$
and $SU(4)$ invariant spinors. Such $N=3$ backgrounds have Killing
spinors
 $\eta_1=f_1 \eta^{SU(5)}$, $\eta_2=f_2 \theta^{SU(5)}$ and
$\eta_3=f_3 \eta^{SU(4)}$.  Combining the conditions of the two classes of $N=2$
backgrounds, we find that
\be
f_1=f_2=f_3=f~,~~~\partial_0f=\partial_5f=\partial_{\bar 5}f=0~.
\la{ssummary00}
\ee
Further combining the condition that the $(3,0)+(0,3)$ parts of $\Omega_{i,jk}$ vanish with the conditions
summarized in section \ref{sumsum}, we find that
the conditions on the $\Omega_{0,0i}$ components are
\be
\Omega_{0,05}=\Omega_{0,0\bar 5}=0~,~~~~
\Omega_{0,0\alpha}=-2\partial_{\a} \log f~.
\la{ssummary1}
\ee
The conditions on the  $\Omega_{0,ij}$ components  are
\be
\Omega_{0, 5\bar\a}=\Omega_{0, 5\a}= \Omega_{0, 5\bar5}=\Omega_{0,\b}{}^\b=0~,~~~~~
\Omega_{0, \b_1\b_2}={i\over4} \Omega_{5,\bar \g_1\bar\g_2} \epsilon^{\bar\g_1\bar\g_2}{}_{\b_1\b_2}
\la{ssummary2}
\ee
and the traceless part of $\Omega_{0, \a\bar\b}$ is not determined.
The conditions on the   $\Omega_{\bar\a, ij}$ components are
\bea
\Omega_{\bar\b_1,\bar\b_2\bar\b_3}&=&0~,~~~
\Omega_{\bar\a, \b_1\b_2}=-\Omega_{0,0[\b_1} g_{\b_2]\bar\a}~,~~~
\Omega_{ \b,\bar\a}{}^\b=-{3\over2} \Omega_{\bar5, \bar\a  5}
=-{3\over 2} \Omega_{0,0\bar\a}~,
\cr
\Omega_{\a,\b}{}^\b&=&-{1\over2}\Omega_{\bar5, \a  5}
=-{1\over2} \Omega_{0,0\a}~.
\la{ssummary3}
\eea
In addition, we have
\bea
&&\Omega_{\bar\b_1, \bar\b_2\bar 5}=\Om_{\bar5,\bar\b_1\bar\b_2}=0~,~~~
\Omega_{[\bar\b_1, \bar\b_2] 5}=-\Omega_{5,\bar\b_1 \bar\b_2}~,~~~
\Omega_{(\bar\b_1, \bar\b_2) 5}=0~,
\cr
&&\Omega_{(\bar\a, \b) \bar5}=\Omega_{(\bar\a, \b) 5}=0~,~~~
\Omega_{\bar\a, 5\bar5}=0~.
\la{ssummary4}
\eea
The traceless part of $\Omega_{\a,\b\bar\g}$ is not determined.
Finally, the conditions on the  $\Omega_{\bar5, ij}$ components are
\bea
&&\Omega_{5,\b}{}^\b=\Omega_{\bar5,\b}{}^\b=0~,~~~
 \Omega_{ 5, \bar\a 5}=\Omega_{\bar 5, \bar\a \bar5}=0~,~~~
\Omega_{ 5, \bar\a \bar5}=\Omega_{\bar 5, \bar\a 5}~,
\cr
&&\Omega_{\bar5, \bar\a  5}= \Omega_{0,0\bar\a}~,~~~
\Omega_{\bar5, 5\bar 5}=\Omega_{5,5\bar 5}=0~.
\la{ssummary5}
\eea
The above equations together with their complex conjugates give full set of conditions that are required for a background
to admit $N=3$ supersymmetry and Killing spinors given by $\eta_1=f_1 \eta^{SU(5)}$, $\eta_2=f_2 \theta^{SU(5)}$ and
$\eta_3=f_3 \eta^{SU(4)}$.

\subsection{Fluxes}

We can substitute the conditions on the geometry that we have
derived into the expressions for the fluxes. As a result, the
electric part of the flux can be written as
\bea
&&G_{\a\b\g}=0~,~~~~G_{5\b\g}=0~,~~~G_{\bar\a 5\g}=-2i \Omega_{\bar\a,5\g}~,
\cr
&&G_{\bar5 \b\g}=-2i \Omega_{\bar5,\b\g}~,~~~G_{\bar\a \b\g}=0~,~~~G_{\bar 5 5\a}=-2i \Omega_{\bar5,5\a}+i
\Omega_{0,0\a}
\eea
Similarly,  the magnetic part of the flux is
\bea
&&F_{\bar\a\b_1\b_2\b_3}=-{3\over2} \Omega_{5,\bar\g_1\bar\g_2}
\epsilon^{\bar\g_1\bar\g_2}{}_{[\b_1\b_2} g_{\b_3]\bar\a}~,~~~
F_{\a_1\a_2\a_3\a_4}=F_{5\a_1\a_2\a_2}=F_{\bar 5\b_1\b_2\b_3}=0~,
\cr
&&
F_{\bar\a 5\b_1\b_2}=0
~,~~~F_{5\bar5 \a}{}^\a=0~,~~~F_{\a\bar\b\g}{}^\g=0~,~~~ F_{\a\b5\bar5}={1\over2}
 \Omega_{5,\bar\g_1\bar\g_2}
\epsilon^{\bar\g_1\bar \g_2}{}_{\a\b}~,~~~
\cr
&& F_{\a\bar\b 5\bar5}=-2i \Omega_{0, \a\bar\b}~,~~~F_{\bar \a\bar5 \b_1\b_2}=0~.
\eea
We have used that the (3,0)+(0,3) part of the connection
$\Omega_{i,jk}$ vanishes. The last two relations are derived from
the conditions for $N=2$ supersymmetry (\ref{solafour}) and
(\ref{solafive}) in appendix D, respectively. The components of the
fluxes that do not appear in the above equations are not determined
by the Killing spinor equations.

\subsection{Geometry}

We shall now investigate some aspects of the spacetime geometry that
arises from the  relations (\ref{ssummary00})-(\ref{ssummary5}).
The geometry of these $N=3$ backgrounds  combines aspects of the geometries
of the $N=2$ backgrounds with $SU(5)$ and $SU(4)$ invariant Killing spinors
 that we have investigated. As in all previous cases, the spacetime admits a timelike
Killing vector field $\kappa^f$ which is inherited from the Killing
spinor $\eta_1$. From the forms associated with the  spinors
$\eta^{SU(5)}$, $\theta^{SU(5)}$ and $\eta^{SU(4)}$, it is clear
that the spacetime admits an $SU(4)$-structure. In particular, the
space $B$  transverse to the orbits of $\kappa^f$ has an
$SU(4)$-structure.  However, unlike the $N=2$ backgrounds with
$SU(4)$ invariant spinors, the space $B$ is  complex and therefore
Hermitian.  This is because the (3,0)+(0,3) parts of the connection
$\Omega_{A,BC}$ vanish. This is similar to the $N=2$ backgrounds
with $SU(5)$ invariant spinors.

There are also two spacelike Killing vector fields $\tilde\kappa^f=if^2
(\partial_{5}-\partial_{\bar5})$ and $\hat\kappa^f=f^2
(\partial_{5}+\partial_{\bar5})$ which are associated with the one-forms (\ref{fvf}) and (\ref{vnff}), respectively,
 constructed
from the spinors $\eta^{SU(5)}$, $\theta^{SU(5)}$ and  $\eta^{SU(4)}$. The Killing vector fields $\kappa^f$ and
$\tilde\kappa^f$ and $\hat\kappa^f$  commute.
The proof of both these statements is similar to those of $N=2$ backgrounds with $SU(4)$ invariant spinors.

The $\tilde \kappa^f$ and $\hat \kappa^f$ preserve the  complex structure of $B$ as well, e.g.
${\cal L}_{\tilde \kappa^f}\omega(\eta^{SU(5)},\eta^{SU(5)})=0$. The computation is similar
to that we presented for the $N=2$ backgrounds with $SU(4)$ invariant Killing spinors.
It is also likely that all three vector fields $\kappa^f$, $\hat \kappa^f$ and $\tilde \kappa^f$
preserve the whole of the $SU(4)$
structure of $B$ including the higher degree forms that are associated with the spinors
$\eta^{SU(5)}$, $\theta^{SU(5)}$ and $\eta^{SU(4)}$.

The space $\hat B$ transverse to the orbits of all three vector
fields is Hermitian with respect to the complex structure associated
with the two-form $\omega^{SU(4)}$, see (\ref{suff}). In addition
$\hat B$ admits an $SU(4)$-structure. We can adapt coordinates to
all the above vector fields and write the metric as
\be
ds^2=U_{ab} (du^a+\b^a) (du^b+\b^b)+\gamma_{IJ} dx^I dx^J
\la{mettem}
\ee
where $U$, $\b$ and $\gamma$ are functions of the $x$ coordinates, and  $a,b=0,1,2$ and $I,J=1,\dots, 8$.

To summarize, we have shown that the $N=3$ backgrounds with  $SU(4)$ invariant
spinors admit  three  commuting Killing vector fields $\kappa^f$, $\hat\kappa^f$
and $\tilde \kappa^f$. The former is timelike while the other two are spacelike. The space $B$ transverse
to the orbits of $\kappa^f$ is a Hermitian manifold with an $SU(4)$-structure and admits two holomorphic Killing
vector fields. The $SU(4)$-structure on $B$ is determined
by the conditions (\ref{ssummary1})-(\ref{ssummary5}), see also appendix B.
Of course these conditions can be put into real form using the  complex structure
and the forms associated with the spinors.

\newsection{$N=4$ backgrounds with $SU(4)$ invariant  spinors}

We can also easily investigate a class of $N=4$ backgrounds, namely
$N=4$ backgrounds that have the Killing spinors $\eta_1=f_1
\eta^{SU(5)}$, $\eta_2=f_2 \theta^{SU(5)}$, $\eta_3=f_3
\eta^{SU(4)}$ and $\eta_4=f_4 \theta^{SU(4)}$, where $f_1,f_2,f_3$ and $f_4$ are real functions
of the spacetime. The conditions coming
from the three first Killing spinors have already been computed in
the previous subsection, and the conditions from $\eta_4$ can be
obtained from the formulas in section \ref{sufcon} by changing
signs on all terms containing the epsilon tensor. For this
compare  the expressions for $\eta^{SU(4)}$ and
$\theta^{SU(4)}$. The conditions we find  from the Killing spinor equations are
\be
f_1=f_2=f_3=f_4=f~,~~~\partial_0f=\partial_{5}f=\partial_{\bar
5}f=0~.
\la{sum0}
\ee
Furthermore, the conditions on the $\Omega_{0,0i}$
components are
\be
\Omega_{0,05}=\Omega_{0,0\bar 5}=-2\partial_5 \log f=-2
\partial_{\bar5} \log f=0~,~~~~ \Omega_{0,0\alpha}=-2\partial_{\a}
\log f~.
\la{sum1}
\ee
The conditions on the  $\Omega_{0,ij}$ components are
\be
\Omega_{0, 5\bar\a}=\Omega_{0, 5\a}= \Omega_{0,
5\bar5}=\Omega_{0,\b}{}^\b=\Omega_{0, \b_1\b_2}=0~,
\la{sum2}
\ee
and the traceless part of $\Omega_{0, \a\bar\b}$ is not determined.
The conditions on the   $\Omega_{\bar\a, ij}$ components are
\bea
\Omega_{\bar\b_1,\bar\b_2\bar\b_3}&=&0~,~~~
\Omega_{\bar\a, \b_1\b_2}=-\Omega_{0,0[\b_1} g_{\b_2]\bar\a}~,~~~
\Omega_{ \b,\bar\a}{}^\b=-{3\over2} \Omega_{\bar5, \bar\a  5}
=-{3\over 2} \Omega_{0,0\bar\a}~,
\cr
\Omega_{\a,\b}{}^\b&=&-{1\over2}\Omega_{\bar5, \a  5}
=-{1\over2} \Omega_{0,0\a}~,
\la{sum3}
\eea
and the traceless part of $\Omega_{\a,\b\bar\g}$ is not determined.
In addition, we have
\bea
&&\Omega_{\bar\b_1, \bar\b_2\bar 5}=\Omega_{\bar\b_1, \bar\b_2
5}=0~,~~~ \Omega_{(\bar\a, \b) \bar5}=\Omega_{(\bar\a, \b) 5}=0~,~~~
\Omega_{\bar\a, 5\bar5}=0~.
\la{sum4}
\eea
Finally, the conditions on the  $\Omega_{\bar5, ij}$ components are
\bea
&&\Omega_{5,\b}{}^\b=\Omega_{\bar5,\b}{}^\b=0~,~~~
 \Omega_{ 5, \bar\a 5}=\Omega_{\bar 5, \bar\a \bar5}=0~,~~~
\Omega_{ 5, \bar\a \bar5}=\Omega_{\bar 5, \bar\a 5}~,
\cr
&&\Omega_{\bar5, \bar\a  5}= \Omega_{0,0\bar\a}~,~~~ \Omega_{\bar5,
5\bar 5}=-\Omega_{5,5\bar
5}=0~,~~~\Om_{5,\bar\a_1\bar\a_2}=\Om_{\bar 5,\bar\a_1\bar\a_2}=0~.
\la{sum5}
\eea
The condition that remains to be examined is (\ref{gtconb}) or equivalently (\ref{gtconbb}).
This gives that
\be
\Omega_{\bar\a,5}{}^{\bar\a}=0~.
\la{sum6}
\ee
The above equations, together with their complex conjugates, give
full set of conditions that are required for a background to admit
$N=4$ supersymmetry with Killing spinors given by $\eta_1=f_1
\eta^{SU(5)}$, $\eta_2=f_2 \theta^{SU(5)}$, $\eta_3=f_3
\eta^{SU(4)}$ and $\eta_4=f_4 \theta^{SU(4)}$.

Using the above conditions,
the fluxes can be expressed in terms of the connection. For the electric part
of the flux, we get
\bea
&&G_{\a\b\g}=G_{5\b\g}=G_{\bar5 \b\g}=G_{\bar\a \b\g}=0~,
\cr
&&G_{\bar\a 5\g}=-2i \Omega_{\bar\a,5\g}~,~~~G_{\bar 5 5\a}=3i
\Omega_{0,0\a}~.
\eea
Similarly, for the magnetic part of the flux we find
\bea
&F_{\a_1\a_2\a_3\a_4}=F_{5\a_1\a_2\a_2}=F_{\bar\a\b_1\b_2\b_3}=F_{\bar
5\b_1\b_2\b_3}=0~,
\cr
& F_{\bar\a 5\b_1\b_2}=F_{5\bar5
\a}{}^\a=F_{\a\bar\b\g}{}^\g=F_{\a\b5\bar5}=0~,
\cr
& F_{\a\bar\b 5\bar5}=-2i \Omega_{0, \a\bar\b}~,~~~F_{\bar \a\bar5
\b_1\b_2}=0~.
\eea
The components of the flux that do not appear in the above equations are not determined by the
Killing spinor equations.
Note that many of the components of the fluxes vanish as a consequence of the requirement
of supersymmetry.

\subsection{Geometry}

We shall now investigate some aspects of the spacetime geometry that
arises from the  relations (\ref{sum0})-(\ref{sum6}). We shall not
elaborate on the description of the geometry  because  the
properties of spacetime in this case are similar to that we have
obtained for $N=2$ and $N=3$ backgrounds. The  spacetime is equipped
an $SU(4)$-structure. This can been seen from the forms associated
with the  spinors $\eta^{SU(5)}$, $\theta^{SU(5)}$, $\eta^{SU(4)}$
and $\theta^{SU(4)}$. In addition, the spacetime admits a timelike
Killing vector field $\kappa^f$ and the ten-dimensional space $B$
transverse to the orbits of this vector field is a Hermitian
manifold. Furthermore, $B$ has two (real) holomorphic vector fields
$\tilde\kappa^f=i f^2 (\partial_5-\partial_{\bar 5})$ and
$\hat\kappa^f=f^2 (\partial_5+\partial_{\bar 5})$. All three vector
fields $\kappa^f, \tilde\kappa^f$ and $\hat\kappa^f$ commute.

The space $\hat B$ transverse to the orbits of all three vector
fields is Hermitian with respect to the complex structure associated
with the two-form $\omega^{SU(4)}$, see (\ref{suff}). In addition
$\hat B$ admits an $SU(4)$-structure. We can adapt coordinates to
all the above vector fields and write the metric as (\ref{mettem}).

To summarize, we have seen that the $N=4$ backgrounds with  $SU(4)$ invariant
spinors admit  three  commuting Killing vector fields $\kappa^f$, $\tilde\kappa^f$
and $\hat \kappa^f$. The first is timelike while the other two  are spacelike. The space $B$ transverse
to the orbits of $\kappa^f$ is a Hermitian manifold with an $SU(4)$-structure and two holomorphic Killing
vector fields. The space $\hat B$ transverse to all three vector fields is also Hermitian
with an $SU(4)$-structure.
The $SU(4)$-structures on  $B$ and $\hat B$ are determined
by the conditions (\ref{sum1})-(\ref{sum6}), see also appendix B.
Of course these conditions can be put into real form using the  complex structure
and the forms associated with these spinors.
Note that the $SU(4)$-structures in the $N=3$ and $N=4$ backgrounds are different.
Some of the components of the spin connection vanish in the latter but they do not vanish in the
former.

\newsection{Calabi-Yau compactifications with fluxes to
one-dimension}

One way to define the Calabi-Yau compactifications of M-theory with
fluxes to one-dimension is to require\footnote{One may in addition
require that the internal manifold is compact.} that the associated
background is invariant under the one-dimensional Poincar\'e group
and that the Killing spinors have stability subgroup $SU(5)$. In the
absence of fluxes, it is clear that such backgrounds become the
standard compactification of M-theory on ten-dimensional Calabi-Yau
manifolds. According to this definition, these Calabi-Yau
compactifications with fluxes are a special class of $N=1$ and $N=2$
supersymmetric backgrounds with $SU(5)$ invariant spinors. In the
absence of fluxes the backgrounds preserve two supersymmetries. In
the presence of fluxes, this is no longer the case. As we shall see
the conditions that arise from the Killing spinor
 equations allow for backgrounds with only one
supersymmetry.

The Poincar\'e group in one-dimension is $\bZ_2\ltimes \bR$. In
particular, the Lorentz group is $\bZ_2$ and the non-trivial element
acts on the one-dimensional Minkowski space as time inversion. In
the context of the backgrounds we are investigating, the Lorentz
group acts as $t\rightarrow -t$ on the coordinate $t$ adapted to the
Killing vector $\kappa^f$ associated with the Killing spinor
$\eta_1$. The  subgroup $\bR$ acts as translations on $t$.

Alternatively, one can require invariance of the background under the connected
Poincar\'e group $\bR$. As we have seen all $N=1$ and $N=2$ backgrounds
with $SU(5)$ invariant spinors admit such a symmetry. Thus,
we have already derived the supersymmetry conditions for such compactifications.
In what follows, we shall focus on the $\bZ_2\ltimes \bR$ case.

\subsection{$N=1$ compactifications}

Requiring invariance under the $\bZ_2$ Lorentz group,
we find that the background can be written as
\bea
ds^2&=&-f^4 dt^2+ ds_{10}^2
\cr
{\rm F}&=&F~.
\eea
In particular the off-diagonal term in the time
component of the metric and the electric part
of the flux vanish. Therefore $e^0=f^2 dt$. Substituting this in
the torsion free condition for the spin connection,
we learn that $\Omega_{i,0j}=\Omega_{0,ij}=0$.
This in particular implies that
\be
F_{\b\bar\a\g}{}^\g=0~.
\ee
Using (\ref{gtco}) and (\ref{gzct}), we find that the vanishing of
the electric flux implies that
\be
\Omega_{[\bar\a_1,\bar\a_2\bar\a_3]}=0~,~~~~~~\Omega_{\bar\a,\b\g}
=g_{\bar\a[\b} \Omega_{0,0\g]}~.
\la{ccone}
\ee
Substituting the first equation in (\ref{ccone}) and $\Omega_{i,0j}=0$
 in (\ref{otF}), we get
\be
F_{\bar\a\b_1\b_2\b_3}={1\over2}
 \Omega_{\bar\a, \bar\g_1\bar\g_2} \epsilon^{\bar\g_1\bar\g_2}{}_{\b_1\b_2\b_3}~.
\ee
Taking the trace of the second equation in (\ref{ccone})
and using  (\ref{gcon}), we find that
\be
\Omega_{\bar\a,}{}^{\bar\a}{}_\g= 2 \Omega_{0,0\g}~,~~~~~
\Omega_{\g,\b}{}^\b=\Omega_{0,0\g}~.
\ee
Therefore, we have
\be
\Omega_{\bar\a,}{}^{\bar\a}{}_\g=2 \Omega_{\g,\b}{}^\b~.
\la{cctwo}
\ee
Substituting this into (\ref{fzF}), we get
\be
F_{\b_1\b_2\b_3\b_4}=-{3\over2} \Omega_{0,0\bar\a}
 \epsilon^{\bar\a}{}_{\b_1\b_2\b_3\b_4}~.
\ee

Therefore one concludes that the Killing spinor equations allow for M-theory Calabi-Yau compactifications
with fluxes to one dimension with one supersymmetry. The non-vanishing fluxes are along the components
$(4,0)+(0,4)$, $(1,3)+(3,1)$ and the traceless part of the $(2,2)$ component of the magnetic flux $F$.
The $(4,0)+(0,4)$ and $(1,3)+(3,1)$ components of $F$ are determined in terms of the ten-dimensional
geometry of $B$. The conditions on geometry of $B$ are given in (\ref{ccone}) and in (\ref{cctwo})~.
The scale factor $f$ is also determined in terms of  the geometry of $B$ as
\be
\Omega_{\g,\b}{}^\b+2\partial_\g f=0~.
\ee

\subsection{$N=2$ compactifications}

The requirement of $N=2$ supersymmetry imposes additional conditions
on the geometry of spacetime for M-theory Calabi-Yau
compactifications with fluxes. As in the $N=1$ case above, the
electric part of the flux and the off-diagonal part of the time
component of the metric vanish as a consequence of the invariance of
the background under the action of $\bZ_2\ltimes \bR$. The latter
condition again implies that $\Omega_{0,ij}=\Omega_{i,0j}=0$. Using
(\ref{tfttct}) and (\ref{tfoct}), we conclude that the $(1,3)+(3,1)$
components and the trace part of the (2,2) component of $F$ vanish.
Thus the only non-vanishing part of $F$ is the traceless part of the
(2,2) component. Since $G=0$,  (\ref{kstecc}) implies that
$\Omega_{0,0\bar\a}=0$ and in turn (\ref{gtcona}) gives that the
scale factor $f$ is constant. In addition $G=0$ and (\ref{tgtcn})
implies that
\be
\Omega_{\bar \a, \b\g}=0~.
\ee
From (\ref{kstmbb}) we also get that
\be
\Om_{\a,\g}{}^\g=0~.
\ee
So the geometry of the spacetime  is simply $\bR\times M_{CY}$,
where $M_{CY}$ is a ten-dimensional Calabi-Yau manifold. Therefore
for M-theory $N=2$ compactifications on Calabi-Yau manifolds with
fluxes, the background is $\bR\times M_{CY}$ and the only
non-vanishing flux allowed is the traceless part of $F^{2,2}$. These
are the conditions required by the Killing spinor equations.

The field equations in both the $N=1$ and $N=2$ cases will impose
additional conditions on the geometry and flux ${\rm F}$. The
Einstein equations arise as integrability conditions of the Killing
spinor equation and therefore
 the independent field equations are those
of the flux \cite{pakis}. The $N=1$ case will be examined elsewhere
\cite{prep}, so we shall focus on the $N=2$ case here. The
supergravity field equations for ${\rm F}$ are $d*{\rm
F}+{1\over2}{\rm F}\wedge {\rm F}=0$. In the $N=2$ case we are
considering ${\rm F}=F$ and the only non-vanishing part is the
traceless part of $F^{(2,2)}$.
  In particular $*F$ has an electric component
and since the spacetime is $\bR\times M_{CY}$, one concludes that
$d*F=0$ and $F\wedge F=0$. Using the traceless condition $F\wedge
F=0$ implies that $|F|^2=0$ and therefore the field equations imply
that the flux vanishes, i.e. ${\rm F}=0$.

This result may change in M-theory because the field equations
 of ${\rm F}$  are modified
by anomaly terms \cite{witten, duffd}. However, it is likely that
the inclusion of the anomaly terms into the theory  changes the
supersymmetry transformations in eleven dimensions  leading
 to a new set of Killing spinor equations,
 see e.g. \cite{green, peeters, howe, gran}. In turn these should be re-investigated and
 the relation between geometry and fluxes may change. There has been
 recent progress in this for a class of M-theory compactifications
 \cite{dragos,chriskellypaul}.

An alternative way to have non-vanishing fluxes in $N=2$
compactifications is to allow ${\rm F}$ to have a non-vanishing
electric component $G$ and thus only require invariance only under
the connected Poincar\'e group $\bR$. An similar analysis has been
done for M-theory compactifications with fluxes on eight-dimensional
Calabi-Yau manifolds \cite{becker}. Such compactifications can also
be reexamined using the conditions for supersymmetry we have derived
for the $N=4$ backgrounds with $SU(4)$ invariant spinors.

\section{Concluding remarks}

We have presented a method to directly solve the Killing spinor
equations of eleven-dimensional supergravity. Our method is based on
the gauge properties of the supercovariant connection and on a
description of spinors in terms of forms. These have led to a better
understanding of the Killing spinors of a supersymmetric M-theory
background and  to a  simplification  of the Killing spinor
equations of eleven-dimensional supergravity. We have given the most
general spinors that can arise in $N=2$ backgrounds and we have
solved the Killing spinor equations for  two  cases associated with
$SU(5)$ and $SU(4)$ invariant spinors\footnote{The most general
$N=2$ case will be presented elsewhere \cite{prep}.}. In the former
case the geometry of spacetime is related to  Hermitian
ten-dimensional  manifolds with an $SU(5)$-structure while in the
latter case the geometry of spacetime is related to ten-dimensional
almost Hermitian  manifolds with an $SU(4)$-structure. In general,
the $G$-structure of a spacetime is related to the stability
subgroup of the Killing spinors in $Spin(1,10)$. We have also
presented two classes of $N=3$ and $N=4$ backgrounds with $SU(4)$
invariant spinors. In both cases the spacetime is related to
Hermitian manifolds with an $SU(4)$-structure which admit
holomorphic vector fields. Our method can be applied to
ten-dimensional supergravities to extend the results of \cite{jfgpa}
to supersymmetric backgrounds with less than maximal supersymmetry.
As an example of backgrounds with $N=1$ and $N=2$ supersymmetry, we
have presented an application to M-theory Calabi-Yau
compactifications with fluxes.

It would be of interest to use  our formalism to express  the
Killing spinors of well-known M-theory backgrounds, e.g. M-branes
\cite{duff, guven} and others, in terms of forms. This may lead to a
better understanding of these solutions of the Killing spinor
equations.  One can  use such  a description
 in terms of forms  to compute
the spacetime forms associated with a pair of Killing spinors. In
many cases such forms are associated with calibrations
\cite{harvey}. In the context of supergravity, they are not closed
and give rise to generalized calibrations \cite{jangp, jangppt}.
The generalized calibrated submanifolds are the  supersymmetric
cycles in these backgrounds, i.e. they are the supersymmetric
solutions of the M-brane worldvolume actions. It is worth pointing
out that in \cite{jangppt} it was observed that the M2- and M5-brane
backgrounds admit such generalized calibrations, see also
\cite{pakis}. These are most likely forms associated to the  Killing
spinors of these backgrounds.

As we have explained the emphasis of our method is in the
description of Killing spinors. We expect that this will also assist
in the physical interpretation of the various supersymmetric
backgrounds that arise as the solutions of the Killing spinor
equations. This is because most supersymmetric backgrounds with
well-known physical interpretation, like e.g. M-branes, have been
found by consideration of the expected isometries of the spacetime
and the number of supersymmetries preserved \cite{duff, guven}. In
addition, one of the criteria used to associate supergravity
backgrounds to dual gauge theories is the number of supersymmetries
that the former and the latter preserve. There may be a way to
refine  this supersymmetry criterion further by finding a direct
interpretation of the stability group of the Killing spinors. We
have already presented such an example in the context of M-theory
Calabi-Yau compactifications with fluxes  to one-dimension. Clearly
the $SU$ series of Killing spinors can be used to extend the
definition of M-theory Calabi-Yau compactifications with fluxes to
other Minkowski spaces. These paradigms can be extended further to
other M-theory compactifications with fluxes. For example,  one can
{\it define} as M-theory $G_2$  compactifications with fluxes as
 those on backgrounds that exhibit the Poincar\'e symmetry of
 four-dimensional Minkowski space, (compact internal space) and admit Killing spinors that have
 stability subgroup $G_2$, see \cite{gukov} for a review and \cite{gppktb} for
 the compactifications without fluxes. A similar characterization
 can be made for M-theory $Spin(7)$ and $Sp(2)$  compactifications with fluxes.

We have seen that the description of $N=2$ backgrounds requires the
investigation of a Gray-Hervella classification of geometric
$G$-structures on a manifold. Some of them have been investigated
already, like for example the standard $SU(5)$-structure in
ten-dimensional Riemannian  manifolds \cite{salamon, pakis,
cabrera}. We have  explored  $SU(n-1)$-structures in
 $2n$-dimensional manifolds in appendix B.
However it is clear from the description of $N=2$ backgrounds that
 many other `exotic' structures
should be investigated. For example, we have seen that some $N=2$
backgrounds exhibit an $SU(2)\times SU(2)$- and an $SU(2)\times
SU(3)$-structure and that many other structures arise as  stability
subgroups of the Killing spinors.

It is clear that our method can be extended to analyze the Killing spinor equations of
supergravity theories in lower dimensions.
This is because the supercovariant derivatives of lower
 dimensional supergravity theories
have a gauge group which includes an appropriate spin group and
there is a description of spinors in terms of forms.  It is expected
that it will be straightforward to carry out our procedure for
lower-dimensional supergravities with a small number of
supersymmetries. This is because in such supergravities the space of
spinors have a low dimension and  the orbits of the spin groups have
been investigated in \cite{bryant}.
 This will provide an independent verification and a
 simplification of results that have
already been obtained, see e.g. \cite{tod, Gauntlett:2002nw},
using other methods. It may also be possible
 to classify  the supersymmetric solutions of lower-dimensional
  supergravities with extended supersymmetry.

\section*{Acknowledgements}

The work of U.G.~is funded by the
Swedish Research Council. This work has been partially supported by the
PPARC grant PPA/G/O/2002/00475.

\setcounter{section}{0}

\appendix{Orbits of special unitary groups}

To understand the Killing spinors of eleven-dimensional supergravity,
one has to find the orbits of $SU(5)$ and $SU(4)$ groups on $\Lambda^2(\bC^5)$ and
$\Lambda^2(\bC^4)$, respectively.
We shall argue here that the generic orbit of $SU(5)$ on
$\Lambda^2(\bC^5)$ is $SU(5)/SU(2)\times SU(2)$ and a representative
complex two-form is $\sigma=\lambda_1 e^1\wedge e^2+\lambda_2
e^3\wedge e^4$, where $\lambda_1, \lambda_2\in\bR$. While the
generic orbit of $SU(4)$ in $\Lambda^2(\bC^4)$ is $SU(4)/SU(2)\times
SU(2)$ and a representative is $\sigma=\lambda_1 e^1\wedge
e^2+\lambda_2 e^3\wedge e^4$, where $\lambda_1, \lambda_2\in \bC$
obeying one real condition, i.e. $|\lambda_1|=1$. We shall produce
two arguments for this. One is based on invariance and the other on
a Lie algebra computation.

The independent $SU(5)$ invariant functions  on $\Lambda^2(\bC^5)$
are
\bea
I_1&=&||\rho||^2=\delta^{a\bar a}
\delta^{b\bar b}\rho_{ab} \bar\rho_{ \bar a \bar b}
\cr
I_2&=&\delta^{b_1\bar a_2} \delta^{a_3\bar b_2} \delta^{b_3\bar a_4}\delta^{a_1\bar b_4}
\rho_{a_1 b_1} \rho_{\bar a_2 \bar b_2}\rho_{a_3 b_3}\rho_{\bar a_4 \bar b_4}
\eea
Both $I_1, I_2$ are real, so it is expected that the generic orbit
has co-dimension two. To see that a representative of the generic
orbit is $\sigma=\lambda_1 e^1\wedge e^2+\lambda_2 e^3\wedge e^4$,
$\lambda_1,\lambda_2\in \bR$, we shall demonstrate that $U^t \sigma
U$, $U\in SU(5)$, is a generic two-form. In fact we shall show that
$\rho=\sigma+i (H^t\sigma+\sigma H)$ is a generic two-form, where
$H$ is a Hermitian traceless matrix. In this way we will show that
 acting with  $SU(5)$ transformations, one can span all the
two-forms at at the linearized level of the action.
Indeed we find that
\begin{small}
\bea
\rho=
\pmatrix{0&\lambda_1 (i+x_1+x_2)&\lambda_1 z_1-\lambda_2 y_3^*&\lambda_1 z_2+\lambda_2y_2^*& \lambda_1 z_3\cr
-\lambda_1 (i+x_1+x_2)& 0&-\lambda_1 y_2-\lambda_2 z_2^*& -\lambda_1 y_3+\lambda_2 z_1^*& -\lambda_1 y_4\cr
-\lambda_1 z_1+\lambda_2 y_3^*&\lambda_1 y_2+\lambda_2 z_2^*&0&\lambda_2 (i+x_4+x_3)& \lambda_2 v_1\cr
-\lambda_1 z_2-\lambda_2y_2^*&-\lambda_2 z_1^*+\lambda_1 y_3&-\lambda_2 (i+x_3+x_4)&0&-\lambda_2 w_2\cr
-\lambda_1 z_3& \lambda_1 y_4&-\lambda_2 v_1& \lambda_2 w_2&0}
\eea
\end{small}
where
\be
H=\pmatrix{x_1& y_1& y_2& y_3& y_4\cr
           y_1^*& x_2& z_1& z_2& z_3\cr
           y^*_2&z_1^*& x_3& w_1& w_2\cr
           y^*_3&z_2^*& w_1^*&x_4& v_1\cr
           y^*_4&z_3^*& w_2^*&v_1^*& x_5}~,~~~~~~x_1+x_2+x_3+x_4+x_5=0~,~~x_1,\dots, x_5\in \bR~.
\ee
It is clear that if the lower diagonal entries of $H$ are generic,
then the lower diagonal entries
of $\rho$ are also generic apart perhaps from $\rho_{21}$ and
$\rho_{43}$ which are determined from
$x_1, \dots, x_5$ and $\lambda_1, \lambda_2$. From the conditions
imposed on $x_1, \dots, x_5$,
$(x_1+x_2)$ is independent from $(x_3+x_4)$ and since $\lambda_1$
is independent from $\lambda_2$,
$\rho_{21}$ and $\rho_{43}$ are also independent and $\rho$ is a generic two form.

As a further confirmation, observe that  the stability
subgroup of $\sigma$ is $SU(2)\times SU(2)$ and so the generic orbit
is $SU(5)/SU(2)\times SU(2)$
which has dimension $18$. On the other hand ${\rm dim}_{\bR} \Lambda^2(\bC^5)=20$. So
$SU(5)/SU(2)\times SU(2)$ has codimension two as expected which is
the number of independent parameters in $\sigma$.

It is clear that there are various special orbits for particular values of $\lambda_1$ and $\lambda_2$. To summarize,
the various orbits are
the following:
\begin{itemize}
\item $\lambda_1\not=\lambda_2\not=0$: stability subgroup $SU(2)\times SU(2)$

\item $\lambda_1=\lambda_2\not=0$: stability subgroup $Sp(2)$

\item $\lambda_1=0$ or $\lambda_2=0$: stability subgroup $SU(2)\times SU(3)$

\item $\lambda_1=\lambda_2=0$: stability subgroup $SU(5)$.

\end{itemize}

The independent $SU(4)$ invariant functions on $\Lambda^2(\bC^4)$ are
\bea
I_1&=&||\rho||^2=\delta^{a\bar a}
\delta^{b\bar b}\rho_{ab} \bar\rho_{ \bar a \bar b}
\cr
I_2&=&\epsilon^{b_1b_2b_2b_4}
\rho_{b_1 b_2} \rho_{b_3b_4}~,
\eea
where $\epsilon$ is the holomorphic volume form. These are three
independent functions and therefore it is expected that the generic
orbit is of co-dimension three. A representative of this orbit is
$\sigma=\lambda_1 e^1\wedge e^2+\lambda_2 e^3\wedge e^4$,
$\lambda_1,\lambda_2\in \bC$. As in the previous case, we shall show
that $U^t \sigma U$, $U\in SU(4)$, spans $\Lambda^2(\bC^4)$ by
computing $\rho=\sigma+i (H^t\sigma+\sigma H)$. Indeed we find that
\begin{small}
\bea
\rho=
\pmatrix{0&\lambda_1 (i+x_1+x_2)&\lambda_1 z_1-\lambda_2 y_3^*&\lambda_1 z_2+\lambda_2y_2^*\cr
-\lambda_1 (i+x_1+x_2)& 0&-\lambda_1 y_2-\lambda_2 z_2^*& -\lambda_1 y_3+\lambda_2 z_1^*\cr
-\lambda_1 z_1+\lambda_2 y_3^*&\lambda_1 y_2+\lambda_2 z_2^*&0&\lambda_2 (i+x_4+x_3)\cr
-\lambda_1 z_2-\lambda_2y_2^*&-\lambda_2 z_1^*+\lambda_1 y_3&-\lambda_2 (i+x_3+x_4)&0}
\eea
\end{small}
where
\be
H=\pmatrix{x_1& y_1& y_2& y_3\cr
           y_1^*& x_2& z_1& z_2\cr
           y^*_2&z_1^*& x_3& w_1\cr
           y^*_3&z_2^*& w_1^*&x_4}~,~~~~~~x_1+x_2+x_3+x_4=0~,~~x_1,\dots, x_4\in \bR~.
\ee
It is clear that if the lower diagonal entries of $H$ are generic, then the lower diagonal entries
of $\rho$ are also generic apart  from $\rho_{21}$ and $\rho_{43}$. These could be
dependent because $x_1+x_2=-x_3-x_4$. However, we can allow for $\lambda_1, \lambda_2\in \bC$.
In such case, $\rho_{21}$ and $\rho_{43}$ are independent. They remain independent after imposing
a condition on $\lambda_1, \lambda_2$, say $|\lambda_1|=1$. Thus we have shown that $\rho$
is a generic two-form.

As a further confirmation, observe that  the stability subgroup of
$\sigma$ is $SU(2)\times SU(2)$ and therefore the generic orbit is
$SU(4)/SU(2)\times SU(2)$ which has dimension $9$. On the other hand
${\rm dim}_{\bR} \Lambda^2(\bC^4)=12$. So $SU(4)/SU(2)\times SU(2)$
has codimension three as expected which is the number of independent
parameters in $\sigma$.

There are various special orbits for particular values of $\lambda_1$ and $\lambda_2$. To summarize,
the various orbits are the following:
\begin{itemize}
\item $\lambda_1\not=\lambda_2\not=0$: stability subgroup $SU(2)\times SU(2)$

\item $\lambda_1=\lambda_2\not=0$: stability subgroup $Sp(2)$

\item $\lambda_1=0$ or $\lambda_2=0$: stability subgroup $SU(2)\times SU(2)$

\item $\lambda_1=\lambda_2=0$: stability subgroup $SU(4)$.

\end{itemize}

\appendix{Geometric $G$-structures}

\subsection{Euclidean signature}

Let  $X$ be a $k$-dimensional manifold equipped with a connection
which takes values in the Lie algebra ${\it g}$ and ${\it h}$ be a
Lie subalgebra ${\it h}\subset so(k)$ and  ${\it h}\subset {\it g}$.
One can classify the compatible ${\it h}$-structures in ${\it g}$ as
follows: Suppose that one can decompose ${\it g}={\it h}\oplus {\it
h}^{\perp}$, where we have assumed that there is a ${\it
g}$-invariant inner product $<,>$ in ${\it g}$. Then the
inequivalent ${\it h}$-structures are labeled by the irreducible
representations of ${\it h}$ in $\bR^k\otimes {\it h}^{\perp}$,
where $\bR^k$ should be thought of as the vector representation of
$so(n)$. This is equivalent to a decomposition of the so called
intrinsic torsion $K$ which is the difference of a ${\it
g}$-connection and a compatible metric ${\it h}$ connection,
$K=\nabla^{{\it g}}-\nabla^{{\it h}}$. In all cases that we
investigate below, there is a unique connection $\nabla^{{\it h}}$
such that $K$ takes values in $\bR^k\otimes {\it h}^{\perp}$
\cite{grayhervella}. The vanishing of one or more components of $K$
in the ${\it h}$-irreducible representation that arise in the
decomposition of $\bR^k\otimes {\it h}^{\perp}$ characterize the
reduction of the ${\it g}$-structure to an ${\it h}$-structure. So
if  $\bR^k\otimes {\it h}^{\perp}$ is decomposed in $r$ ${\it
h}$-irreducible representations, then there are $2^r$ inequivalent
compatible ${\it h}$ reductions of the ${\it g}$-structure. Then we
say that the manifold admits an ${\it h}$-structure or an
H-structure. In what follows we shall not be concerned with global
topological issues which cannot be addressed from a local
description of the supergravity solutions. Instead, we shall use the
H-structures to characterize the geometry of supersymmetric
backgrounds.

{}For applications to Riemannian manifolds, ${\it g}\subseteq so(k)$
and $\nabla^{{\it g}}$ is the Levi-Civita connection $\nabla$. If
the ${\it h}$-structure is characterized by the presence of
invariant tensors, which we denote collectively with $\alpha$, then
the components of the intrinsic torsion
 can be represented by $\nabla\alpha$.
In particular, this  applies to the supersymmetric
 backgrounds  which admit   Killing
 spinors that have a non-trivial stability
subgroup ${\it h}$. This is because the associated spacetime forms $\alpha$
constructed from the spinors are ${\it h}$ invariant.
However as we shall explain below, we can identify the
components of the intrinsic torsion
 from those of the spin connection  using the frame that
 comes naturally with the formalism we have developed
 in this paper. Of course this new way of identifying
the components of the intrinsic torsion
is equivalent to $\nabla\alpha$ and the two are related by a linear transformation.

{}For the Gray-Hervella classification of almost Hermitian manifolds
 one takes ${\it g}=so(2n)$, $so(k)=so(2n)$, ${\it h}=u(n)$, $n\geq 3$
 and $\nabla^{so(2n)}$ to be  the Levi-Civita connection.
It is known that $\bR^{2n}\otimes u(n)^{\perp}$ is decomposed into
four irreducible representations under $u(n)$ and therefore there
are sixteen almost Hermitian
 types of manifolds compatible with an $so(2n)$-structure.
In the cases where we have an adapted $u(n)$ frame,
 as in the case of supersymmetric backgrounds
that we are investigating, the intrinsic torsion can be represented
 by the components of the $so(2n)$ spin-connection
\be
\Omega_{\bar\a, \b\g}~,~~~~\Omega_{\bar\a, \bar\b\bar\g}~
\la{cotor}
\ee
and their complex conjugates. To compare this  with the standard definition
of the intrinsic torsion
as $\nabla \omega$, where $\omega$ is the K\"ahler form, we find that
\bea
\nabla_{\bar \a}\omega_{\b\g}&=&2i \Omega_{\bar\a,\b\g}
\cr
\nabla_{\bar\a}\omega_{\bar\b\bar\g}&=&-2i \Omega_{\bar\a,\bar\b\bar\g}~.
\la{sit}
\eea
The components $\Omega_{\bar\a, \b\bar\g}$
of the connection and their complex conjugates take values in $u(n)$.
It suffices to  decompose   the (\ref{cotor})
components of the intrinsic  torsion under the four
 irreducible $u(n)$ representations in $\bR^{2n}\otimes u(n)^{\perp}$
 because the remaining components are not independent
and they are related to the above by complex conjugation. The first
component of the intrinsic  torsion can be decomposed under $su(n)$
in a trace and a traceless part, i.e.
\be
(\ww_3)_{\g}=\Omega_{\bar\b,}{}^{\bar\b}{}_\g~,~~~~~ (\ww_4)_{\bar\a \b\g}
=\Omega_{\bar\a, \b\g}-{2\over n-1}
\Omega_{\bar\d,}{}^{\bar\d}{}_{[\g} g_{\b]\bar\a}
\ee
and the second component as
\be
(\ww_2)_{\bar\a \bar\b\bar\g}=\Omega_{[\bar\a, \bar\b\bar\g]}~,~~~~~~~~
(\ww_1)_{\bar\a \bar\b\bar\g}=
{2\over 3}\Omega_{\bar\a, \bar\b\bar\g}-{1\over 3}\Omega_{\bar\g, \bar\a\bar\b}
-{1\over3} \Omega_{\bar\b, \bar\g\bar\a}~.
\ee
The vanishing of one or more of the above four components of the
intrinsic  torsion characterize the sixteen classes
of almost Hermitian manifolds. Using (\ref{sit}) one can relate
the $\ww$ classes to the ${\cal W}$ classes of the Gray-Hervella
classification \cite{grayhervella}. In particular, the
 $\ww_3$ class is related to the ${\cal W}_3$ class and similarly for the other classes.

In the above formalism it is straightforward to extend
the classification of $u(n)$-structures to the $su(n)$-structures, $n\geq3$
\cite{salamon, cabrera}.
In the latter case the independent components of the intrinsic  torsion are
\be
\Omega_{\bar\a, \b\g}~,~~~~\Omega_{\bar\a, \bar\b\bar\g}~, ~~~~~~~~
\Omega_{\bar\a,\b}{}^\b~.
\la{cotorsu}
\ee
The space $\bR^{2n}\otimes su(n)^{\perp}$ decomposes into five
irreducible representations under $su(n)$. Therefore there are thirty two
inequivalent $su(n)$-structures compatible with an $so(2n)$-structure.
The decomposition of the intrinsic  torsion under $su(n)$ is
\bea
&&(\ww_3)_\a=\Omega_{\bar\b,}{}^{\bar\b}{}_\a~,~~~~~(\ww_4)_{\bar\a
\b\g}= \Omega_{\bar\a, \b\g}-{2\over n-1}
\Omega_{\bar\d,}{}^{\bar\d}{}_{[\g} g_{\b]\bar\a}
~,~~~~~(\ww_5)_{\bar\a}=\Omega_{\bar\a,\b}{}^\b~,
\cr
&&(\ww_1)_{\bar\a \bar\b\bar\g}=\Omega_{[\bar\a, \bar\b\bar\g]}~,~~~~~~~~
(\ww_2)_{\bar\a \bar\b\bar\g}={2\over 3}
\Omega_{\bar\a, \bar\b\bar\g}-{1\over 3}\Omega_{\bar\g, \bar\a\bar\b}
-{1\over3} \Omega_{\bar\b, \bar\g\bar\a}~.
\eea
The vanishing of one or more of the above five components of the
intrinsic  torsion characterize the thirty two classes
of almost Hermitian manifolds with an $su(n)$-structure.

Now let us turn to investigate the $u(n-1)$-structures of an
$so(2n)$ manifold. For this we split the index $\alpha=(i, n)$,
where $i=1,\dots,n-1$, $n\geq 4$. (The analysis below works for
$n\leq 4$ as well but some of the classes vanish identically.) The
independent components of the intrinsic  torsion in this case are
\bea
&&\Omega_{\bar i, jk}~,~~~~\Omega_{\bar i, \bar j\bar k}~,~~~~~~
\Omega_{\bar n, jk}~,~~~~\Omega_{\bar n, \bar j\bar k}~,~~~~
\Omega_{\bar i, n\bar n}~,~~~~\Omega_{\bar n, n\bar n}~,
\cr
&&\Omega_{\bar i, nj}~,~~~\Omega_{\bar i, \bar n j}~,~~~\Omega_{\bar i, n \bar j}~,~~~\Omega_{\bar i, \bar n \bar j}~,~~~
\Omega_{\bar n, nj}~,~~~\Omega_{\bar n, \bar n j}~,~~~\Omega_{\bar n, n \bar j}~,
~~~\Omega_{\bar n, \bar n \bar j}~.
\la{bsev}
\eea
The space $\bR^{2n}\otimes u(n-1)^{\perp}$ decomposes into 20 irreducible
representations under $u(n-1)$. Therefore there are $2^{20}$ inequivalent compatible
$u(n-1)$-structures in an $so(2n)$ manifold. The intrinsic  torsion decomposes under $u(n-1)$ as
\bea
&&(\ww_3)_k=\Omega_{\bar j,}{}^{\bar j}{}_k~,~~~~~
(\ww_4)_{\bar ijk}=\Omega_{\bar i, j k}-{2\over (n-2)}  \Omega_{\bar m}{}^{\bar m}{}_{[k} g_{j]\bar i}
~,
\cr
&&(\ww_5)_{jk}=\Omega_{\bar n, jk}~,~~~~(\ww_6)_{\bar j\bar k}=\Omega_{\bar n, \bar j\bar k}~,
\cr
&&(\ww_1)_{\bar i\bar j\bar k}=\Omega_{[\bar i, \bar j\bar k]}~,~~~~~(\ww_2)_{\bar i\bar j\bar k}={2\over 3}\Omega_{\bar i, \bar j\bar k}
-{1\over 3}\Omega_{\bar k, \bar i\bar j}
-{1\over3} \Omega_{\bar j, \bar k\bar i}~,
\cr
&&(\ww_7)_{\bar i}=\Omega_{\bar i, n\bar n}~,~~~~~~~\ww_8=\Omega_{\bar n, n\bar n}~,
\eea
and another twelve. The first eight arise from taking traces and
traceless parts of $\Omega_{\bar i, nj}$ and $\Omega_{\bar i, \bar n
j} $,  and symmetrizing and skew-symmetrizing $\Omega_{\bar i, n
\bar j}$ and $\Omega_{\bar i, \bar n \bar j}$. The last four
components in (\ref{bsev}) are irreducible. The vanishing of one or more of the
above components of the intrinsic  torsion characterizes the
$2^{20}$ inequivalent $u(n-1)$-structures of an $so(2n)$ manifold.

To investigate the $su(n-1)$-structures of an $so(2n)$ manifold,
we again split the index $\alpha=(i, n)$, where $i=1,\dots,n-1$.
The independent components of the intrinsic  torsion in this case are
\bea
&&\Omega_{\bar i, jk}~,~~~~\Omega_{\bar i, \bar j\bar k}~,~~~~~~
\Omega_{\bar n, jk}~,~~~~\Omega_{\bar n, \bar j\bar k}~,~~~~
\Omega_{\bar i, n\bar n}~,~~~~\Omega_{\bar n, n\bar n}~,~~~~~~
\Omega_{\bar i, j}{}^j~,~~~~~~
\Omega_{\bar n, j}{}^j~,
\cr
&&\Omega_{\bar i, nj}~,~~~\Omega_{\bar i, \bar n j}~,~~~\Omega_{\bar i, n \bar j}~,~~~\Omega_{\bar i, \bar n \bar j}~,~~~
\Omega_{\bar n, nj}~,~~~\Omega_{\bar n, \bar n j}~,~~~\Omega_{\bar n, n \bar j}~,~~~\Omega_{\bar n, \bar n \bar j}~.
\la{bnine}
\eea
The space $\bR^{2n}\otimes su(n-1)^{\perp}$ decomposes into 22 irreducible
representations under $su(n-1)$. Therefore there are $2^{22}$ inequivalent compatible
$su(n-1)$-structures in an $so(2n)$ manifold. The intrinsic  torsion
decomposes under $su(n-1)$ as
\bea
&&(\ww_3)_k=\Omega_{\bar j,}{}^{\bar j}{}_k~,~~~~~(\ww_4)_{\bar ijk}=
\Omega_{\bar i, j k}-{2\over (n-2)}  \Omega_{\bar m}{}^{\bar m}{}_{[k} g_{j]\bar i}
~,
\cr
&&(\ww_5)_{jk}=\Omega_{\bar n, jk}~,~~~~(\ww_6)_{\bar j\bar k}=\Omega_{\bar n, \bar j\bar k}~,~~~~~~
(\ww_9)_{\bar i}=\Omega_{\bar i, j}{}^j~,
\cr
&&(\ww_1)_{\bar i\bar j\bar k}=\Omega_{[\bar i, \bar j\bar k]}~,~~~~~~~~(\ww_2)_{\bar i\bar j\bar k}
={2\over 3}\Omega_{\bar i, \bar j\bar k}
-{1\over 3}\Omega_{\bar k, \bar i\bar j}
-{1\over3} \Omega_{\bar j, \bar k\bar i}~,~~~~
\cr
&&
(\ww_7)_{\bar i}=\Omega_{\bar i, n\bar n}~,~~~~\ww_8=\Omega_{\bar n, n\bar n}~,~~~~~~
(\ww_{10})=\Omega_{\bar n, j}{}^j~,
\eea
and another twelve. The first eight arise from taking traces and
traceless parts of $\Omega_{\bar i, nj}$ and $\Omega_{\bar i, \bar n
j} $,  and symmetrizing and skew-symmetrizing $\Omega_{\bar i, n
\bar j}$ and $\Omega_{\bar i, \bar n \bar j}$. The last four
components in (\ref{bnine}) are irreducible. The vanishing of one or more of the
above components of the intrinsic  torsion characterizes the
$2^{22}$ inequivalent $su(n-1)$-structures of an $so(2n)$ manifold.

\subsection{Lorentzian signature}

So far we have investigated ${\it g}$-structures for Euclidean
signature manifolds. The analysis can be easily extended to the
Lorentzian signature manifolds. We shall not examine Lorentzian case
in detail since it is similar to the Euclidean case. Instead, we
shall present an example of the $so(n)$-structures in an $so(n,1)$
manifold. This case is relevant in the context of $N=1$
supersymmetric backgrounds. The independent components of the
intrinsic  torsion in this case are
\be
\Omega_{i,j0}~,~~~\Omega_{0,0 i}~,
\ee
where we have split the frame index $A=(0,i)$, $i=1,\dots,n$.
The space $\bR^{n+1}\otimes so(n)^\perp$ decomposes under $so(n)$
into four irreducible representations. Therefore there are sixteen
inequivalent  $so(n)$-structures compatible with an $so(n,1)$-structure.
The intrinsic  torsion decomposes under $so(n)$ as
\bea
({\it w}_1)_{ij}=\Omega_{[i,j]0}~,~~~{\it w}_2=\Omega_{i,}{}^i{}_0~,~~~({\it w}_3)_{ij}=\Omega_{(i,j)0}-
{1\over n} g_{ij}\Omega_{k,}{}^k{}_0~,~~~({\it w}_4)_i=\Omega_{0,0 i}~.
\la{sons}
\eea
The vanishing of one or more of the above components of the intrinsic  torsion characterizes
the sixteen inequivalent $so(n)$-structures of an $so(n,1)$ manifold.
As we have seen the vanishing of the second and the third class
in (\ref{sons}) is related to the existence of a time-like Killing vector
on the Lorentzian manifold.

We can combine the results of this section with those we have presented for
 ${\it g}$-structures
on Euclidean signature manifolds. In particular,
we can find the $u(n)$-, $su(n)$-, $u(n-1)$- and
$su(n-1)$-structures of an $so(2n,1)$ manifold. For example it is easy to see
that there are 256
$u(n)$-structures in an $so(2n,1)$ manifold.
The conditions
for $N=1$ and $N=2$ supersymmetry,
that we have derived in this paper, can be viewed as
 particular $su(5)$- and $su(4)$- structures
in an $so(10,1)$ manifold. On the other hand, it is equivalent
 to express the various
conditions arising from the Killing spinor equations  in
terms of the spacetime connection
as we have done in the most part of the paper. The two ways
 of expressing the conditions
of the spacetime geometry are related by a linear transformation.

\appendix{Spinors with stability subgroup $(Spin(7)\ltimes \bR^8)\times \bR$}

To find a representative
for the orbit of $Spin(1,10)$ with stability subgroup
$(Spin(7)\ltimes \bR^8)\times \bR$, it is sufficient to
find a spinor associated to a null vector.
We first consider the spinor
\be
a e_1+b e_{2345}~,~~~~~~~a,b\in \bC~.
\ee
The Majorana condition implies that
\be
a=\bar b~.
\ee
Therefore, we construct two real spinors
\bea
&&e_1+e_{2345}
\cr
&&i(e_1-e_{2345})~.
\eea
Using these, we write
\bea
\eta^{Spin(7)}&=&{1\over  2} (i(1-e_{12345})+e_1+e_{2345})~.
\eea
Next we compute the associated vector to find
\bea
\kappa_0(\eta^{Spin(7)}, \eta^{Spin(7)})&=& B(\eta^{Spin(7)}, \Gamma_0 \eta^{Spin(7)})=-1
\cr
\kappa_1(\eta^{Spin(7)}, \eta^{Spin(7)})&=& B(\eta^{Spin(7)}, \Gamma_1 \eta^{Spin(7)})=1
\eea
with the rest of the components to vanish, i.e. $\kappa=-e^0+e^1$.
Clearly this is null and  $\eta^{Spin(7)}$ is a representative
of the  orbit ${\cal O}_{Spin(7)}$.

\appendix{The solutions of the Killing spinor equations for $SU(4)$ invariant spinors}

To solve the Killing spinor equations for the $SU(4)$
 invariant spinor, we substitute the expression for the fluxes we have derived
 in the $N=1$ case. Then we use the resulting
 equations to determine  the remaining components of the fluxes
 and find the independent  components of the spin connection.
Throughout
 this calculation, we  use the condition
$\Omega_{i,0j}=\Omega_{0,ij}$ which arises from the Killing spinor
equations for $\eta_1$ and (\ref{gcon}) which expresses
 $\Omega_{0,0i}$  in terms of the
geometry of the ten-dimensional manifold $B$.

First we investigate the solutions of the Killing spinor equations
that involve a derivative along the frame time direction. Using
(\ref{fzF}) and (\ref{tG}), (\ref{tfone}) and its complex conjugate
imply that
\be
\Omega_{0,0\bar5}=\Omega_{0,05}
\la{soloone}
\ee
and
\be
-2 \Omega_{0,05}+ (\Omega_{5,5\bar5}-\Omega_{\bar5,5\bar5})
+(\Omega_{5,\b}{}^\b-\Omega_{\bar5,\b}{}^\b)=0~.
\la{solotwo}
\ee
Next we observe that using (\ref{ttzF}) and (\ref{gzct}),
(\ref{tftwo}) implies that
\be
4 \Omega_{0, \bar \a 5}+i \Omega_{\b_1,\b_2\b_3} \epsilon^{\b_1 \b_2\b_3}{}_{\bar\a}=0~.
\la{solothree}
\ee
The equation (\ref{tfthree}) and its complex conjugate imply that
\be
\partial_0\log g=0
\la{solofour}
\ee
and
\be
\Omega_{0,\b}{}^\b-\Omega_{0,5\bar 5}+{i\over12} F_\a{}^\a{}_\b{}^\b
-{i\over 6} F_\a{}^\a{}_{5\bar5}=0~.
\la{soloo}
\ee
The latter using (\ref{ttzF}) can be rewritten as
\be
\Omega_{0,5\bar5}=\Om_{0,\g}{}^\g~.
\la{solofive}
\ee

Eliminating the fluxes in (\ref{tffour}) using  (\ref{otF}), we find
\be
i \Omega_{5, \bar\b_1\bar\b_2}+ i \Omega_{[\bar\b_1, \bar\b_2] \bar5}
-{1\over2} \Omega_{0, \g_1\g_2} \epsilon^{\g_1\g_2}{}_{\bar\b_1\bar\b_2}=0~.
\la{solosix}
\ee
Similarly using (\ref{gtco}) and (\ref{otF}), we get from (\ref{tffive})  that
\be
2\Omega_{\bar 5, \bar
5\bar\a}+2\Omega_{5,\bar\a\bar5}-5\Omega_{0,0\bar\a}-2\Omega_{\b,\bar\a}{}^{\b}=0~.
\la{soloseven}
\ee

Next we turn to investigate the condition that
 are associated with the Killing spinor equations involving
the spatial derivative along the $\bar \alpha$ frame direction. It
is easy to see, using (\ref{ttzF}) and (\ref{gzct}), that
(\ref{afone}) gives (\ref{solothree}) and so (\ref{afone}) is not
independent. Next, substituting  (\ref{gtco}) and (\ref{otF}) into
(\ref{aftwo}), we find that
\bea
\Omega_{\bar\a, \bar\b5}-\Omega_{\bar\a, \bar\b\bar5} +{1\over3}
\Omega_{5,\bar\a\bar\b}-\Omega_{\bar 5,\bar\a\bar\b} -{2\over3}
\Omega_{[\bar\a,\bar\b]\bar5} -{i\over3} \Omega_{0,\g_1\g_2}
\epsilon^{\g_1\g_2}{}_{\bar\a\bar\b}=0~.
\la{solaone}
\eea
Eliminating the fluxes from (\ref{afthree}) using (\ref{otF}) and
(\ref{gtco}), we get
\bea
\partial_{\bar\a}\log g+{1\over2} \Omega_{\bar\a,\g}{}^\g-{1\over2}\Omega_{\bar\a,5\bar5}-{1\over6} \Omega_{\b,\bar\a}{}^\b
+{1\over6}\Omega_{5,\bar\a\bar5}+{2\over3}\Omega_{\bar5, \bar5\bar\a}-
{1\over6}\Omega_{0,0\bar\a}=0~.
\la{solatwo}
\eea
Similarly using (\ref{otF}) and (\ref{gtco}), (\ref{affour}) gives
\bea
\Omega_{\bar\a,\g_1\g_2}\epsilon^{\g_1\g_2}{}_{\bar\b_1\bar\b_2}
+[{1\over3}\Omega_{5,5\d}-{1\over3}\Omega_{\bar\b,\d}{}^{\bar\b}+{1\over3}\Omega_{\bar5,\d 5}
+{1\over6}\Omega_{0,0\d}]\epsilon^\d{}_{\bar\a\bar\b_1\bar\b_2}=0~.
\la{solathree}
\eea
Next we find using (\ref{otF}) and (\ref{gtco})  that the condition
(\ref{affive}) gives (\ref{solosix}). As can be seen using
(\ref{tF}) and (\ref{ttzF}), we cannot eliminate all the components
of the a fluxes from
 (\ref{afsix}) and we get
\bea
i \Omega_{\bar\a, 0\g}+{1\over2}
F_{\bar\a\g5\bar5}-\frac{i}{3}\Omega_{0,\b}{}^\b g_{\bar\a\g}-
\frac{2i}{3} \Omega_{0,5\bar5} g_{\bar\a\g}=0~.
\la{solafour}
\eea
Similarly using  (\ref{gtco}) and (\ref{ttzF}), (\ref{afseven}) gives
\bea
{1\over4} F_{\bar\a\bar5\g_1\g_2}-{1\over8}
(\Omega_{\bar\a,\bar\b_1\bar\b_2}
+\Omega_{[\bar\a,\bar\b_1\bar\b_2]}) \epsilon^{\bar\b_1\bar\b_2}{}_{\g_1\g_2}
+{i\over3} g_{\bar\a[\g_1} \Omega_{\bar5, 0\g_2]}=0~.
\la{solafive}
\eea
The condition (\ref{afeight}) relates the partial derivative of $g$
to the geometry of spacetime.
 Using (\ref{gtco}), one finds that
\bea
\partial_{\bar\a}\log g-{1\over2} \Omega_{\bar\a,\g}{}^\g
-{1\over2} \Omega_{\g,\bar\a}{}^\g
+{1\over2} \Omega_{\bar\a,5\bar5}
+{1\over2}\Omega_{5,\bar\a\bar5}-{1\over2}\Omega_{0,0\bar\a}=0~.
\la{solasix}
\eea
Eliminating the fluxes from (\ref{afnine}) using (\ref{gtco}) and (\ref{fzF}),
we find
\bea
{1\over3}[-{1\over2}\Omega_{0,05}-{1\over2}\Omega_{0,0\bar5}-\Omega_{5,\g}{}^\g
+\Omega_{\bar5,\g}{}^\g-\Omega_{5,5\bar5}+\Omega_{\bar5,5\bar5}] g_{\bar\a\b}
+\Omega_{\bar\a,\b\bar5}+\Omega_{\b,\bar\a\bar5}=0~.
\la{solaseven}
\eea
Similarly using (\ref{otF}), we get from (\ref{aften})  that
\bea
4i\Omega_{\bar\a,0\bar5}-\Omega_{\g_1,\g_2\g_3} \epsilon^{\g_1\g_2\g_3}{}_{\bar\a}=0~.
\la{solaeight}
\eea

Next we turn to investigate the conditions that are associated with
the Killing spinor equations involving the spatial derivative along
the $\bar 5$ frame direction. As can be seen using (\ref{tF}) and
(\ref{ttzF}), the condition (\ref{fffone}) gives \be
2\Om_{0,5\bar5}+\Om_{0,\g}{}^\g=0
\ee and together with (\ref{solofive}) imply
\be \Om_{0,5\bar5}=\Om_{0,\g}{}^\g=0~.
\ee

Next we use (\ref{gtco}) and (\ref{otF}) to write (\ref{ffftwo}) as
\bea
\Omega_{\bar5,\bar\b5}-{2\over3} \Omega_{5,\bar\b\bar5}+{1\over3}
\Omega_{\bar5,\bar5\bar\b} -{1\over3}
\Omega_{\a,\bar\b}{}^\a-{5\over6}\Omega_{0,0\bar\b}=0~.
\la{solfone}
\eea
Using (\ref{gtco}), we can express
the partial derivative of $g$, (\ref{fffthree}), as
\bea
\partial_{\bar5} \log g-\Omega_{\bar5,5\bar5}-{1\over2}\Omega_{0,0\bar5}=0~.
\la{solftwo}
\eea
Eliminating the fluxes from (\ref{ffffour}) using (\ref{gtco}) and (\ref{otF}),
we find  that
\bea
{1\over2} \Omega_{5,\g_1\g_2}\epsilon^{\g_1\g_2}{}_{\bar\b_1\bar\b_2}
+{5\over6}\Omega_{\g_1,\g_25} \epsilon^{\g_1\g_2}{}_{\bar\b_1\bar\b_2}
+{1\over3} \Omega_{\bar5, \g_1\g_2} \epsilon^{\g_1\g_2}{}_{\bar\b_1\bar\b_2}
-{4i\over3} \Omega_{0,\bar\b_1\bar\b_2}=0~.
\la{solfthree}
\eea
As we shall see (\ref{solfthree}) is not an independent condition.
Similarly using  (\ref{otF}), we  find that condition (\ref{fffive})
gives (\ref{solaeight}) and so it is not independent. It is
straightforward to see using (\ref{gzct}) and (\ref{ttzF}) that
(\ref{fffsix}) implies  (\ref{solothree}) and therefore it is not
independent. Next we find using (\ref{gzct}) that the condition
(\ref{fffseven}) gives
\bea
\Omega_{5, \b_1\b_2}+\Omega_{[\b_1,\b_2]5}=0~.
\la{solffour}
\eea
One can show
using (\ref{fzF}) and (\ref{gzct}) that (\ref{fffeight}) implies
\bea
\partial_{\bar5}\log g-{2\over3} (\Omega_{\bar5,\g}{}^\g-\Omega_{5,\g}{}^\g)
+{1\over3} \Omega_{\bar 5, 5\bar5}+{2\over3} \Omega_{5, 5\bar5}
-{1\over6} \Omega_{0,0\bar5}+{1\over3} \Omega_{0,05}=0~,
\la{solffive}
\eea
and similarly
using (\ref{fzF}) that (\ref{fffnine}) gives
\bea
\Omega_{0,0\a}+2\Omega_{\a,5\bar5}+2\Omega_{\a,\g}{}^\g+2 \Omega_{\bar5,\a\bar5}=0~.
\la{solfsix}
\eea
This concludes the substitution of the fluxes in terms of the
geometry in  the conditions that arise from the Killing spinor
equations for $\eta_2$.

The conditions that we have derived involving the  connection
 can be interpreted as restrictions on the geometry of spacetime.
 These can be solved to find the independent components of the connection.
For this we  use (\ref{gcon}) which expresses $\Omega_{0,0i}$ in terms of the
geometry of $B$. In particular,
the component $\Omega_{0,05}$ of the connection can be expressed
in terms of the geometry of $B$. As a result
(\ref{soloone}) and (\ref{solotwo}) can be written as
\be
(\Omega_{\bar\b}{}^{\bar\b}{}_5-\Omega_{\b}{}^{\b}{}_{\bar5})- (\Omega_{5,\b}{}^\b+\Omega_{\bar5,\b}{}^\b)
- (\Omega_{5,5\bar5}+\Omega_{\bar5,5\bar5})=0
\la{anonea}
\ee
and
\be
-(\Omega_{\bar\b}{}^{\bar\b}{}_5+\Omega_{\b}{}^{\b}{}_{\bar5})
+ 2 (\Omega_{5,\b}{}^\b-\Omega_{\bar5,\b}{}^\b)
+2 (\Omega_{5,5\bar5}-\Omega_{\bar5,5\bar5})=0~,
\la{anoneb}
\ee
respectively. Alternatively, we can use (\ref{gconb}) to find that (\ref{soloone})
implies that
\be
\partial_{5}f-\partial_{\bar 5} f=0~.
\la{anone}
\ee
The condition (\ref{solothree}) can be solved in
terms of $\Omega_{0, \bar\a 5}$ to find
\be \Omega_{0, \bar\a 5}=
-{i\over 4} \Omega_{\b_1,\b_2\b_3}
\epsilon^{\b_1\b_2\b_3}{}_{\bar\a}~,
\la{antwo}
\ee
which together with (\ref{solaeight}) implies
\be \Om_{0,\bar\alpha
5}=\Om_{0,\bar\alpha \bar 5}~.
\la{anthree}
\ee
The condition (\ref{solofour}) implies that $g$ is independent of
the frame time direction. The condition (\ref{solosix}) can be solved to reveal that
\be
\Omega_{0,\b_1\b_2}={i\over4} (\Omega_{5,\bar\g_1\bar\g_2}+
\Omega_{\bar\g_1,\bar\g_2 \bar 5})
\epsilon^{\bar\g_1\bar\g_2}{}_{\b_1\b_2}~.
\la{anfour}
\ee
The condition (\ref{soloseven}) restricts the geometry of the space
$B$ as can be easily seen using (\ref{gcon}). Substituting
(\ref{anfour}) into (\ref{solaone}), we find
\be
\Omega_{\bar\a,\bar\b 5}-\Omega_{\bar\a,\bar\b
\bar5}+\Omega_{5,\bar\a\bar\b} -\Omega_{\bar5,\bar\a\bar\b}=0~.
\la{anfive}
\ee
In particular, this gives
\be \Omega_{(\bar\a,\bar\b)
5}-\Omega_{(\bar\a,\bar\b) \bar5}=0
\la{ansix}
\ee and
\be
\Omega_{[\bar\a,\bar\b] 5}+\Omega_{5,\bar\a\bar\b}=0~,
\la{anseven}
\ee
where in the last step we have used (\ref{solffour}).
The condition (\ref{solathree}) will be examined later.
The condition (\ref{solafour}) determines the flux $F_{\bar\a\g
5\bar 5}$ in terms of the connection.
The condition (\ref{solafive}) expresses $F_{\bar\a\bar 5\g_1\g_2}$
in terms of the connection and by taking the trace and comparing to
(\ref{otF}) we find
\be
F_{\g\bar 5\d}{}^\d=\Om_{0,\g\bar 5}=0~.
\la{annine}
\ee
This in turn implies, using (\ref{antwo}), that the totally
anti-symmetric part of the connection vanishes, i.e.
$\Om_{[\b_1,\b_2\b_3]}=0$.
By adding (\ref{solatwo}) and (\ref{solasix}), and using
(\ref{soloseven}), we get
\be
\partial_{\bar\a}g=\partial_{\bar\a}f~.
\la{anten}
\ee
The difference between (\ref{solatwo}) and (\ref{solasix}), after using (\ref{soloseven}), is
\be
\Omega_{\bar\a,\b}{}^\b-\Omega_{\bar\a, 5\bar5}+\Omega_{\bar5,\bar5\bar\a}-{1\over2} \Omega_{0,0\bar\a}=0~.
\la{antenb}
\ee
The condition (\ref{solaseven}) together with (\ref{solotwo})
implies that
\be
\Om_{(\bar\a,\b),\bar 5}=\frac{1}{2}g_{\bar \a\b}\Om_{0,0\bar 5}~.
\la{aneleven}
\ee
By also taking (\ref{soloone}) into account we find
\be
\Om_{(\bar\a,\b)5}=\Om_{(\bar\a,\b)\bar 5}~.
\la{antwelve}
\ee
Taking the trace of the two relations above yields
\bea
&&\Om_{\bar\g,}{}^{\bar\g}{}_{\bar 5}+\Om_{\g,}{}^\g{}_{\bar 5}=4\Om_{0,0\bar 5}~,\la{anthirteena}\\
&&\Om_{\bar\g,}{}^{\bar\g}{}_{ 5}+\Om_{\g,}{}^\g{}_{5}=\Om_{\bar\g,}{}^{\bar\g}{}_{\bar 5}+\Om_{\g,}{}^\g{}_{\bar 5}~.
\la{anthirteenb}
\eea
By adding (\ref{solftwo}) and (\ref{solffive}), we find
\be
\partial_{\bar 5}\log g+\frac{1}{3}(\Om_{5,\g}{}^\g-\Om_{\bar 5\g}{}^\g)+\frac{1}{3}(\Om_{5,5\bar 5}-\Om_{\bar 5,5\bar
5})-{1\over6} \Omega_{0,05}=0~.
\la{anfourteen}
\ee
Together with its complex conjugate this equation gives
\be
(\partial_{\bar 5}-\partial_5)g=0~.
\la{anfifteen}
\ee
If we instead subtract (\ref{solffive}) and (\ref{solftwo}), we get
\be
-(\Om_{\bar 5,\g}{}^\g-\Om_{5,\g}{}^\g)+2\Om_{\bar 5,5\bar
5}+\Om_{5,5\bar 5}+\Om_{0,05}=0~,
\la{ansixteen}
\ee
where we have also used (\ref{soloone}). Taking the complex conjugate of this equation, we find
\be
\Om_{5,5\bar 5}+\Om_{\bar 5,5\bar 5}=0~,
\la{anseventeen}
\ee
and
\be
-(\Om_{\bar 5,\g}{}^\g-\Om_{5,\g}{}^\g)+\Omega_{\bar5,5\bar5}+\Omega_{0,05}=0~.
\la{anseventeenb}
\ee
Comparing the above equation with (\ref{solotwo}), we find
\be
\Om_{0,05}=\Om_{5,5\bar 5}
\la{anseventeenc}
\ee
and
\be
\Om_{5,\g}{}^\g=\Om_{\bar 5\g}{}^\g~.
\ee
Substituting (\ref{anseventeenc}) into (\ref{solftwo}), we get
\be
\partial_{\bar 5}g=\partial_{\bar 5}f~,
\la{antenc}
\ee
where we have also used (\ref{gconb}).
Therefore (\ref{anten}) and (\ref{antenc}) imply that
\be
f=g~.
\ee
The equation (\ref{solfthree}) is not independent. One can see this by
using (\ref{solffour}) and by comparing with  (\ref{anfour}).

Let us now turn to investigate (\ref{solathree}). This can be
written as
\be
\Omega_{\bar\a, \g_1\g_2}+[{1\over3} \Omega_{5,5[\g_1}-{1\over3} \Omega_{\bar\b,[\g_1}{}^{\bar\b}-
{1\over 3} \Omega_{\bar 5,5[\g_1} +{1\over6} \Omega_{0,0[\g_1}] g_{\g_2]\bar\a}=0~.
\la{solathreeb}
\ee
Taking the trace, we get
\be
\Omega_{\bar \b,\a}{}^{\bar\b}+ \Omega_{5,5\a}+\Omega_{\bar 5,\a
5}+{1\over2} \Omega_{0,0\a}=0~.
\la{solathreetr}
\ee

It remains to investigate the equations, (\ref{solathreetr}),
(\ref{solfsix}), (\ref{antenb}),  (\ref{solfone}) and
(\ref{soloseven}). By adding (\ref{solfsix}) and (\ref{antenb}) we
find
\be
2\Omega_{\a, 5\bar5}+\Omega_{\bar 5,\a \bar 5}+\Omega_{5,5\a}=0
\la{twobi}
\ee
and by subtracting we get
\be
2\Omega_{\a, \b}{}^\b+ \Omega_{0,0\a}+\Omega_{\bar 5, \a\bar 5}-\Omega_{5,5\a}=0~.
\la{twostar}
\ee
Eliminating $\Omega_{\bar \b,\a}{}^{\bar\b}$ and $\Omega_{\a, \b}{}^\b$ using
(\ref{solathreetr}) and (\ref{twostar}) from the $\a$-component of (\ref{gcon}),
we find
\be
\Omega_{5,5\a}+\Omega_{\bar 5, \a\bar5}-2\Omega_{\a, 5\bar5}=0~.
\la{sixa}
\ee
Comparing this with (\ref{twobi}), we get
\be
\Omega_{\a, 5\bar5}=0~,~~~~~~\Omega_{5,5\a}+\Omega_{\bar 5,\a\bar5}=0~.
\la{sixb}
\ee
Eliminating $\Omega_{\bar \b,\a}{}^{\bar\b}$  using
(\ref{solathreetr}) from  (\ref{solfone}) and (\ref{soloseven}), we get
\be
\Omega_{5,\a\bar5}-{1\over3}\Omega_{\bar 5,\a5}+{2\over3} \Omega_{5,5\a}-{2\over3}\Omega_{0,0\a}=0
\ee
and
\be
\Omega_{5,5\a}+\Omega_{\bar 5,\a 5}-\Omega_{0,0\a}=0~.
\la{sixc}
\ee
The former using the latter becomes
\be
\Omega_{5,\a\bar5}+\Omega_{5,5\a}-\Omega_{0,0\a}=0~.
\la{sevena}
\ee
Subtracting (\ref{sixc}) from (\ref{sevena}),   we find
that
\be
\Omega_{5,\a\bar5}=\Omega_{\bar5,\a5}~.
\la{sevenb}
\ee
Substituting (\ref{sevena}) into (\ref{solathreetr}), and using (\ref{sevenb}), we get
\be
\Omega_{\bar\b,\a}{}^{\bar\b}=-{3\over2} \Omega_{0,0\a}~.
\la{sevenc}
\ee
In turn substituting (\ref{sevena}) and (\ref{sevenc})  into (\ref{solathreeb}), we find that
\be
\Omega_{\bar\a, \b_1\b_2}+\Omega_{0,0[\b_1} g_{\b_2]\bar\a}=0~.
\ee
We also find from (\ref{twostar}) that
\be
2\Omega_{\a,\b}{}^\b+\Omega_{\bar5,\a\bar5}+\Omega_{\bar5,\a5}=0~.
\ee
This concludes our analysis. The final results are summarized in
section (\ref{sumsum}).

\end{document}